\documentclass[aps,prb,twocolumn,amsmath,amssymb,showpacs]{revtex4}
\usepackage[linktocpage=true]{hyperref}
\usepackage{graphicx}
\usepackage{booktabs}
\usepackage{color}
\usepackage{epsfig}

\begin{document}
   \title{
   Interference and multi-particle effects in a Mach-Zehnder interferometer with single-particle sources
         } 
         \author{Guillem Rossell\'o$^{1}$, Francesca Battista$^{1,2}$, Michael Moskalets$^{3}$, and Janine Splettstoesser$^{1,4}$}
         \affiliation{ $^1$Institut f\"ur Theorie der Statistischen Physik, RWTH Aachen University, D-52056 Aachen, Germany
          \& JARA-Fundamentals of Future Information Technology\\
         $^2$Departamento de F\'isica, FCEyN, Universidad de Buenos Aires and IFIBA,
Pabell\'on I, Ciudad Universitaria, 1428 CABA Argentina \\
$^{3}$Department of Metal and Semiconductor Physics, NTU ``Kharkiv Polytechnic Institute", 61002 Kharkiv, Ukraine\\
 $^4$Department of Microtechnology and Nanoscience (MC2), Chalmers University of Technology, SE-41298 G\"oteborg, Sweden
}        
\date\today


\begin{abstract}
We investigate a Mach-Zehnder interferometer fed by two time-dependently driven single-particle sources, one of them placed in front of the interferometer, the other in the centre of one of the arms.  As long as the two sources are operated independently, the signal at the output of the interferometer shows an interference pattern, which we analyse in the spectral current, in  the charge and energy currents, as well as in the charge current noise. The synchronisation of the two sources in this specifically designed setup allows for collisions and absorptions of particles at different points of the interferometer, which have a strong impact on the detected signals. It introduces further relevant time-scales and can even lead to a full suppression of the interference in some of the discussed quantities. The complementary interpretations of this phenomenon in terms of  spectral properties and tuneable two-particle effects (absorptions and quantum exchange effects) are put forward in this article.
 \end{abstract}

\maketitle

\section{Introduction}

The coherent emission of single particles into a nano-electronic circuit can be realised by the time-dependent modulation of mesoscopic structures. Recently, the creation of Lorentzian current pulses carrying exactly one electron charge,~\cite{Dubois13,Levitov:1996ie,Ivanov:1997wz} the realisation of periodically driven mesoscopic capacitors as single-particle sources by time-dependent gating,~\cite{Feve07,Moskalets08,Bocquillon12} the emission of particles from quantum dots with surface-acoustic waves~\cite{McNeil11,Hermelin11,Wanner14}, as well as particle emission from dynamical quantum dots~\cite{Blumenthal07,Kaestner08,Fletcher13,Ubbelohde15} have been intensively studied. 
Nano-electronic devices fed by these single-particle sources allow for the observation of controlled and tuneable quantum-interference and multiple-particle effects and even for the combination of both.~\cite{Keeling06,Keeling08,Olkhovskaya08,Splett08,Splett09,Moskalets09,Splett10,Mahe10,Juergens11,Moskalets11,Haack11,Bocquillon12,Jonckheere12,Bocquillon13,Haack13,Ferraro13}

\begin{figure}[b]
\includegraphics[width=3.4in]{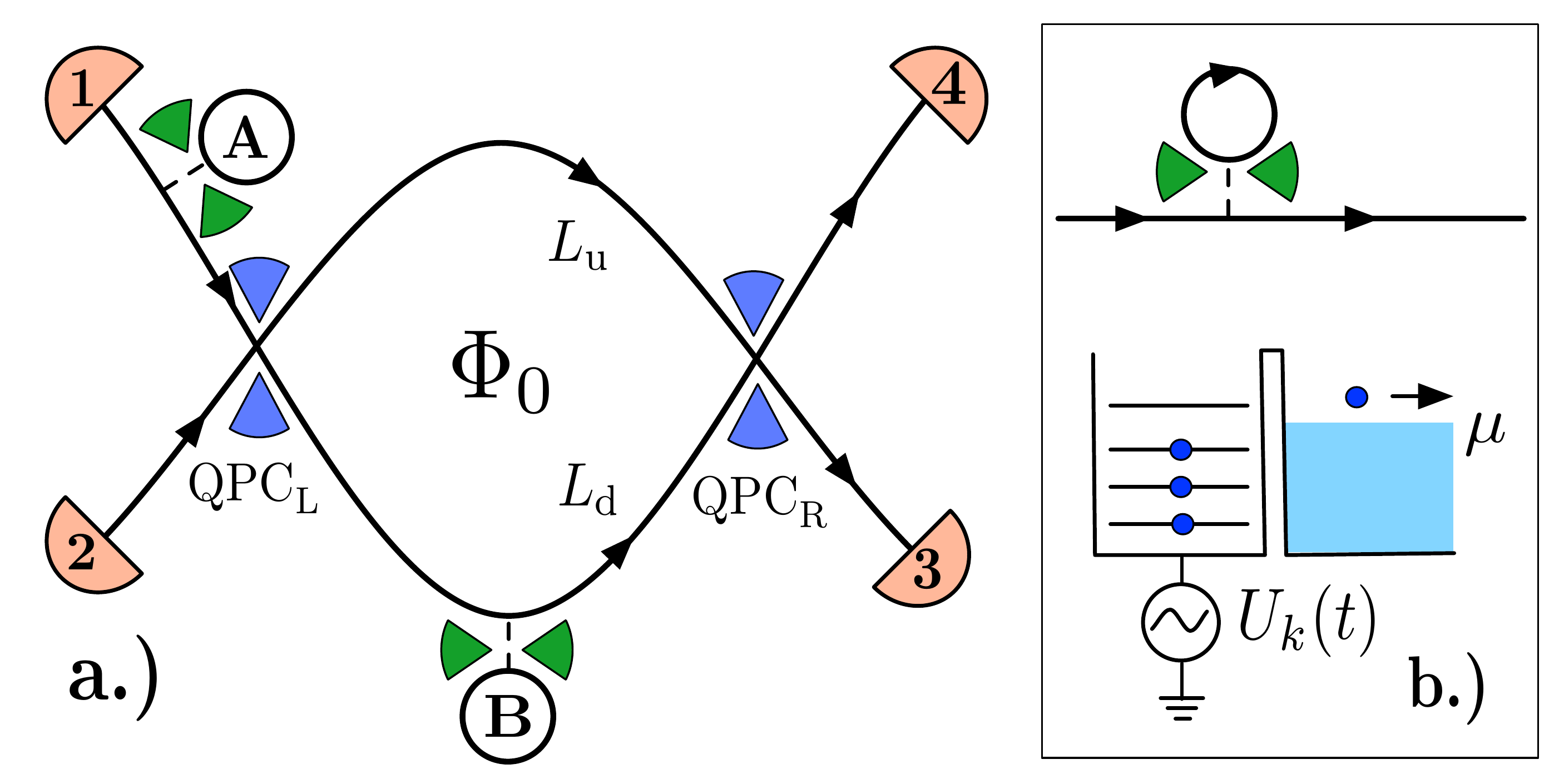}
\caption{a.) Schematic of an electronic MZI. Transport takes place along edge states (black lines; arrows indicate their chirality). Quantum point contacts, QPC$_\text{L}$ and QPC$_\text{R}$, act as beam splitters. All the reservoirs are grounded and particles are injected into the system by two single-particle sources SPS$_\text{A}$ and SPS$_\text{B}$. Charge and energy currents are measured at reservoir~4.
b.) Schematic of an SPS, which is realised by a mesoscopic capacitor. It is implemented as a circular edge state and periodically driven by a potential $U_k(t)$, emitting one electron and one hole per period. }
\label{fig_setup}
\end{figure}

A useful tool to observe quantum-interference effects in an electronic system is a Mach-Zehnder interferometer (MZI),~\cite{Ji03,Litvin07,Neder07,Roulleau07,Huynh12,Yamamoto12,Weisz14,Bautze14} as sketched in Fig.~\ref{fig_setup}~a.), which can be realised by edge states in Quantum Hall systems with the help of quantum point contacts (QPCs). It has been shown that the investigation of the output current of an MZI, when fed by a single-particle source (SPS), such as the one realised by F\`eve \textit{et al.},~\cite{Feve07} see also Fig.~\ref{fig_setup}~b.), allows for the extraction of an electronic \textit{single-particle coherence time}. More generally, it carries interesting new features of coherence properties of the travelling particles.~\cite{Gaury14,Moskalets14}
The combination of several of these sources makes it possible to study controlled two-particle effects, for example the electronic analogue of the Hong-Ou-Mandel effect,~\cite{Olkhovskaya08,Jonckheere12,Iyoda14,Khan15} which was realised experimentally by Bocquillon \textit{et al.}~\cite{Bocquillon12} and Dubois \textit{et al.}~\cite{Dubois13} The combination of several MZIs and SPSs is a possibility to create and detect time-bin entanglement.~\cite{Splett09,Vyshnevyy12,Vyshnevyy13}
However, the impact of controlled multiple-particle effects on the interference pattern detected in electronic interferometers was studied only sparsely~\cite{Juergens11,Hofer14} and leaves a number of open questions concerning the interplay of the two effects. 

In this work, we investigate an MZI into which particles are injected  from an SPS, such that quantum interference effects can be detected at the interferometer output.  The signal detected at the output shows intriguing features due to the energy-dependent transmission of the MZI. Subsequently, a second SPS is introduced injecting particles after the first SPS. Particularly interesting is the case when the second SPS injects particles into one of the interferometer arms, only.  The setup is chosen such that two-particle effects, namely the collision and absorption of particles,~\cite{Splett08} can be observed in different parts of the interferometer. 
We use this setup to carefully investigate the occurrence of tuneable two-particle effects from synchronised SPSs in an electronic MZI, as shown in Fig.~\ref{fig_setup}a.). The particle emission (and absorption) from the second source has a tuneable impact on the interference effects obtained from the signal of the first SPS.  
In order to visualize this impact, we study the spectral properties of the detected signal, the charge and energy currents,~\cite{Butcher90} as well as the charge-current noise,~\cite{Blanter00} based on a Floquet scattering-matrix approach.~\cite{Moskalets02} We here neglect Coulomb interaction, which can lead to relaxation and decoherence~\cite{Ferraro14} of the injected single particles and which is expected to modify our results at most on a quantitative level.~\cite{Wahl14}

Importantly, the observables that we investigate theoretically in this paper, can be envisaged to be studied also in experiments. Indeed, the charge current and charge-current noise of SPSs in Quantum Hall devices was recently measured.~\cite{Mahe10,Parmentier12,Bocquillon12,Bocquillon13} Measurements of the spectral current in the stationary regime in edge states out of equilibrium have been presented in Ref.~\onlinecite{Altimiras10}. Also energy-resolved currents of time-dependently driven single-electron sources were measured~\cite{Ubbelohde15,Fletcher13} and give access to the spectral current as well as to the energy current. Measurements of interference effects in energy or heat currents via changes in the reservoir temperature were detected in a stationary superconducting interferometer.~\cite{Giazotto12} 

The theoretical study presented here, investigates in detail the effect of {\it a coherent suppression of interference} appearing when the two SPSs are properly synchronized. This effect allows for two complementary types of interpretation, related to the \textit{spectral properties} and to the \textit{particle nature} of the injected signals. The spectral current  gives an insight into the behaviour of plane waves as the constituents of the complex signal of the MZI with one or two sources. The reason for this is that the spectral current yields information on the \textit{energy-resolved} interference pattern. With the knowledge on these spectral properties we can explain the features occurring in the \textit{energy-integrated} charge and energy currents.
At the same time, we show that it is in certain cases useful to explain the suppression of interference in the charge and energy current by the occurrence of \textit{two-particle effects}: the placement of SPS$_\text{B}$ in the lower arm of the interferometer introduces the possibility of tuneable particle collisions and absorptions permitting to distinguish the paths traversed by the particles (which-path information).
In order to reliably investigate the impact of two-particle effects (namely through absorption and quantum exchange) we analyse the charge-current noise, obtained from a correlation function of two current operators, which is hence able to capture two-particle physics directly.

The paper is organised as follows. We introduce the system and the investigated observables, as well as the scattering matrix approach employed by us in Sec.~\ref{sec_technical}.  The presentation of results starts with the spectral current, the charge and the energy current for the case of an interferometer fed by one SPS only, in Sec.~\ref{sec_single}. In Sec.~\ref{sec_synchronized}, this is followed by a study of the same quantities in an MZI where particles from two SPSs can collide or where particles can get absorbed. Finally results for the charge-current noise are shown in Sec.~\ref{sec_noise}. In the Appendix, all relevant analytic results which are not presented explicitly in the main text are summarised.

\section{Model and Technique}\label{sec_technical}

\subsection{Mach-Zehnder interferometer with two single-particle sources}\label{sec_model}

The electronic analogue of an MZI, as sketched in Fig. \ref{fig_setup}a.), can be  realised in a two-dimensional electron gas in the quantum Hall regime.~\cite{Ji03,Roulleau07,Litvin07} In these setups, transport takes place along spin-polarised, chiral edge states depicted as black lines in Fig.~\ref{fig_setup}~a.), where arrows indicate their chirality. Two quantum point contacts, QPC$_\ell$, $\ell=\text{L}, \text{R}$, with energy-independent transmission (reflection) amplitudes $t_\ell$ ($r_\ell$) and the related transmission (reflection) probabilities $T_\ell=|t_\ell|^2$ ($R_\ell=|r_\ell|^2$) act as beam splitters. The incoming electronic signal is reflected or transmitted at QPC$_\text{L}$, into the upper arm (u) or the lower arm (d) of the interferometer, with the respective length $L_\text{u}$ and $L_\text{d}$. At QPC$_\text{R}$ the signal is finally reflected or transmitted into reservoir~3 or  4. Assuming a linear dispersion with the drift velocity $v_\text{D}$, the traversal time of the interferometer arms is given by $\tau_\text{u}=L_\text{u}/v_\text{D}$ and $\tau_\text{d}=L_\text{d}/v_\text{D}$.
The interferometer is penetrated by a magnetic flux $\Phi_0$. Therefore, the phase acquired by the electronic wave function due to the propagation along the upper and the lower arm is given by  $\phi_\text{u/d}=\Phi_\text{u/d}+E\tau_\text{u/d}/\hbar$ with the energy-dependent dynamical phase $E\tau_\text{u/d}/\hbar$ and the energy-independent part,  $\Phi_\text{u/d}$, including the magnetic-flux  contribution $\Phi_0$.  The energy and charge currents observed at the detector are known to depend on the difference between the two phases,  $\Delta\phi(E, \Phi)=\Phi+E\Delta\tau/\hbar$ with $\Phi=\Phi_\text{u}-\Phi_\text{d}$ and  the  detuning, $\Delta\tau=\tau_\text{u}-\tau_\text{d}$ of  the traversal times of the interferometer, which is a measure of the imbalance  of the interferometer. 
We assume the extensions of the MZI to be smaller than the dephasing length, which can be limited due to environment- and interaction-induced effects.~\cite{Chalker07,Neuenhahn08,Roulleau08,Roulleau09,Bieri09}
The electronic reservoirs, $\alpha=1,2,3,4$, are at temperature $\theta$ and they are grounded at the equilibrium chemical potential $\mu$, which we take as the zero of energy from here on. 

Particles - electrons and holes - are injected into the MZI by means of a controllable single-particle source, SPS$_\text{A}$, situated at the channel incoming from reservoir~1. A second single-particle emitter, SPS$_\text{B}$, is placed at the lower arm at $L_\text{d}/2$. 
We take the SPSs to be mesoscopic capacitors which are time-dependently driven by periodic gate potentials as sketched in  Fig.~\ref{fig_setup}~b.), inspired by the  experimental realisation by F\`eve {\it et al.}~\cite{Feve07} These SPS$_k$, with $k=$A,B, consist of a quantum dot with a discrete spectrum, weakly coupled  to an edge state through a QPC$_k$. A periodically oscillating time-dependent gate voltage $U_k(t)$, with period $\mathcal{T}=2\pi/\Omega$ and frequency $\Omega$, moves the energy levels of the respective quantum dot, such that one of the levels is subsequently driven above and below the electro-chemical potential $\mu$. This triggers the emission of an electron from source $k=\text{A,B}$ at time $t_k^\text{e}$, during one half of the driving period, and the emission of a hole (which is equivalent to the absorption of an electron) at a time $t_k^\text{h}$ during the other half of the period. 

This particle emission from SPS$_k$ leads to current pulses carrying one electron or one hole. The injection of current pulses from SPS$_\text{A}$  into the MZI, results in an interference pattern in the detected observables at the output of the interferometer.~\cite{Haack11,Haack13} This is in contrast to the current pulses emitted from SPS$_\text{B}$ which travel along the lower arm only and therefore do not create an interference pattern on their own. 

However, the \textit{synchronisation of the two sources}, obtained by tuning the phase difference between the two driving potentials $U_k(t)$, influences the interference pattern drastically.~\cite{Juergens11} The synchronisation of the two sources results in collisions of particles (i.e. the overlap of current pulses carrying an electron each, respectively carrying a hole each) at SPS$_\text{B}$ or QPC$_\text{R}$ or in an absorption process (i.e. the overlap of a current pulse carrying an electron with a current pulse carrying a hole) at SPS$_\text{B}$. It has been shown in Ref.~\onlinecite{Juergens11} that these collisions and absorptions add a non-trivial phase to the interference pattern in the \textit{time-resolved current} at the detector at the output of the MZI, which can even lead to the full suppression of interference in the detected \textit{average charge current}. 
Of particular relevance for these synchronised two-particle events are the two time-differences $\Delta t^{ij}_\text{d}$, $\Delta t^{ij}_\text{u}$. The first one is the difference between the time at which a particle $i=$e,h emitted from SPS$_\text{A}$ travelling the lower arm arrives at SPS$_\text{B}$ and the emission time of a particle $j=$e,h at SPS$_\text{B}$, $\Delta t^{ij}_\text{d}\equiv t_\text{A}^i-t_\text{B}^j+\tau_\text{d}/2$. The second one is the difference between the time at which a particle $i$ emitted from SPS$_\text{A}$ travelling the upper arm arrives at QPC$_\text{R}$ and the time at which a particle $j$ emitted from SPS$_\text{B}$ arrives at QPC$_\text{R}$, $\Delta t^{ij}_\text{u}\equiv t_\text{A}^i-t_\text{B}^j+\tau_\text{u}-\tau_\text{d}/2$. 

\subsection{Scattering matrix formalism}

We describe the transport properties of the above introduced  system  with the help of a Floquet scattering matrix formalism.  Due to the time-periodic modulation of the SPSs, coherent \textit{inelastic scattering} can take place. Thus the scattering matrix elements $S_{\alpha\beta}(E_n,E_m)$, connect the incoming currents from reservoir $\beta$ at energy $E_m=E+m\hbar\Omega$ to the outgoing currents at reservoir $\alpha$ at energy $E_n=E+n\hbar\Omega$ differing from the incoming energy by an integer multiple $n-m$ of the energy quantum $\hbar\Omega$ given by the driving frequency (Floquet quanta).~\cite{Moskalets02} These scattering matrices can be conveniently written in terms of the partial Fourier transforms,
\begin{subequations}
\begin{eqnarray}\label{deffloq}
S_{\alpha\beta}(E_n,E_m)&=&\int_0^{\mathcal{T}}\frac{dt}{\mathcal{T}} e^{i(n-m)\Omega t}S_{\text{in,}\alpha\beta}(t,E_m)\\
S_{\alpha\beta}(E_n,E_m)&=&\int_0^{\mathcal{T}}\frac{dt}{\mathcal{T}} e^{-i(n-m)\Omega t}S_{\text{out,}\alpha\beta}(E_n,t) .\ \ 
\end{eqnarray}
\end{subequations}
Here, $S_{\text{in,}\alpha\beta}(t,E_m)$ is the dynamical scattering amplitude  for a current signal incoming from reservoir $\beta$ at energy $E_m$ to be detected at a time $t$ at reservoir $\alpha$, while $S_{\text{out,}\alpha\beta}(E_n,t)$ is the dynamical scattering matrix for a current signal incoming from reservoir $\beta$ at time $t$ to be found at energy $E_n$  at reservoir $\alpha$.~\cite{Moskalets08}

In this work, we are interested in the regime of \textit{adiabatic driving}, namely where the dwell time of a particle in the mesoscopic capacitor constituting the SPS  is much smaller than the modulation period $\mathcal{T}$ of the driving potential.~\cite{Splett08} Note that this is an assumption on the time-scales describing the SPSs and their driving only, and does not concern the time-scales describing the traversal of the interferometer which can be of arbitrary magnitude. The result is that time-dependent current pulses of Lorentzian shape are emitted into the MZI. This is similar to the recently realised "levitons",~\cite{Dubois13} which are of Lorentzian shape as well.	
In the adiabatic regime, the dynamical scattering matrices describing the subsystem of an SPS, $S_k(t)$ for $k=\text{A,B}$, are energy independent on the scale of the driving frequency and $S_{\text{in,}k}(t,E)=S_{\text{in,}k}(t,\mu)=S_{\text{out,}k}(E,t)=S_{\text{out,}k}(\mu,t)\equiv S_k(t)$. For weak coupling and slow driving of the sources,  these scattering matrices are given by,~\cite{Olkhovskaya08}
\begin{equation}\label{eq_adiabatic_S}
S_k(t)=n_k^\text{e}\frac{t-t_k^\text{e}+i\sigma_k}{t-t_k^\text{e}-i\sigma_k}+n_k^\text{h}\frac{t-t_k^\text{h}-i\sigma_k}{t-t_k^\text{h}+i\sigma_k}\ .
\end{equation}
The emission times of electrons and holes, $t_k^i$, and the width of the emitted current pulses,  $\sigma_k$, are directly related to the properties of the sources and are thus tuneable.~\cite{Splett08} We introduced the variables $n_k^i$ in order to distinguish whether the emission of an electron or of a hole is treated.  This variable takes the value $n_k^\text{e/h}=1$ if a time-interval where an electron/hole is emitted from source $k$ is considered, and $n_k^\text{e/h}=0$ otherwise. We assume that electron and hole emission happen at times, which differ from each other by much more than the pulse width $\sigma_k$, $|t^\text{e}_k-t^\text{h}_k|\gg\sigma_k$, meaning  that the different current pulses emitted from the same source are well separated. 
The scattering matrices of the full system including the MZI and SPSs are given in Appendix~\ref{app_Smatrix}.

\subsection{Observables}

In this paper, we study the impact of two-particle effects on the flux-dependence of the charge current, the energy current, and their spectral functions, as well as on the zero-frequency charge-current noise. In this section we introduce the studied observables.

We start from the time-resolved charge~\cite{Buttiker:1992vr} and energy~\cite{Sergi:2011eo,Battista:2013ew,Ludovico:2014de} current operators in lead $\alpha$, $\hat{I}_{\alpha}(t)$ and $\hat{J}_{\alpha}(t)$, defined as 
\begin{eqnarray}
\hat{I}_{\alpha}(t)&=&\frac{-e}{h}\int^{\infty}_{-\infty} dE \int^{\infty}_{-\infty} dE' e^{i(E-E')t/\hbar}\ \hat{i}_{\alpha}(E,E')\label{eq_defcurr}\\
\hat{J}_{\alpha}(t)&=&\frac{1}{h}\int^{\infty}_{-\infty} dE \int^{\infty}_{-\infty} dE' e^{i(E-E')t/\hbar}\nonumber\\
&&\times\left[\frac{(E+E')}{2}\right]\hat{i}_{\alpha}(E,E')\label{eq_defencurr}\ .
\end{eqnarray}
Note that in this setup the energy current with respect to the electrochemical potential $\mu$ equals the heat
current, since no voltages or temperature gradients are applied. Here, we introduced the operator $\hat{i}_{\alpha}(E,E')=[\hat{b}^{\dagger}_{\alpha}(E)\hat{b}_{\alpha}(E')-\hat{a}^{\dagger}_{\alpha}(E)\hat{a}_{\alpha}(E')]$, and the electron charge $-e$.
The creation and annihilation operators, $\hat{b}^{\dagger}_{\alpha}(E)$ and $\hat{b}^{}_\alpha(E)$, of particles incident in reservoir $\alpha$ are related to the respective operators for particles emitted from reservoir $\beta$ onto the scattering region, $\hat{a}^{\dagger}_{\beta}(E)$ and $\hat{a}^{}_\beta(E)$,  through the Floquet scattering matrix introduced in the previous section by  
\begin{equation}\label{eq_abrel}
\hat{b}^{\dagger}_{\alpha}(E)=\sum_{\beta}\sum_{n=-\infty}^\infty S^*_{\alpha\beta}(E,E_n)\hat{a}^{\dagger}_{\beta}(E_n),
\end{equation}
(and equivalently for the annihilation operators). 

We are interested in the time-averaged charge and energy currents, $\bar{I}_\alpha$ and $\bar{J}_\alpha$, which are given by the time integral over the expectation values of Eqs.~(\ref{eq_defcurr}) and~(\ref{eq_defencurr}),
\begin{eqnarray}
\bar{I}_{\alpha}&=&  \int_0^\mathcal{T} \frac{dt}{\mathcal{T}} \langle \hat{I}_{\alpha}(t) \rangle\label{eq_defaverage1}\\ 
\bar{J}_{\alpha}&=&  \int_0^\mathcal{T} \frac{dt}{\mathcal{T}} \langle \hat{J}_{\alpha}(t) \rangle\label{eq_defaverage2}\ .
 \end{eqnarray}
Here, $\langle\dots\rangle$ indicates a quantum-statistical average. The quantum-statistical average of  particles incoming from the reservoirs is given by the  Fermi function $f(E)=[1+\text{exp}(E/k_\text{B}\theta)]^{-1}$, namely the equilibrium distribution function of the reservoirs, $\langle\hat{a}^{\dagger}_{\alpha}(E)\hat{a}_{\alpha}(E')\rangle=f(E)\delta(E-E')$.
Substituting Eq.~(\ref{eq_abrel}) into Eqs.~(\ref{eq_defcurr}) and~(\ref{eq_defencurr}) and taking the time-average of the expectation values as given in Eqs. (\ref{eq_defaverage1}) and (\ref{eq_defaverage2}), we find 
\begin{eqnarray}
\bar{I}_{\alpha} & = & \frac{-e}{h}\int^{\infty}_{-\infty} dE \ i_{\alpha}(E)\label{eq_current_density}\\
\bar{J}_{\alpha} & = &\frac{1}{h} \int^{\infty}_{-\infty} dE \ E \ i_{\alpha}(E)\label{eq_energy_density} .
\end{eqnarray}
The excess-energy distribution function $i_{\alpha}(E)$, which we also refer to as the spectral current, entering the two current expressions is given by~\cite{Moskalets02,Altimiras10}
\begin{eqnarray}\label{eq_energy_distribution}
i_{\alpha}(E) & = &\int_0^\mathcal{T} \frac{dt}{\mathcal{T}} \int^{\infty}_{-\infty} dE' e^{i(E-E')t/\hbar}\ \langle \hat{i}_\alpha(E,E')\rangle\nonumber\\
& = & \sum_{\beta}\sum_{n=-\infty}^\infty |S_{\alpha\beta}(E,E_n)|^2[f(E_n)-f(E)]\ .
\end{eqnarray}
It describes the distribution of electron and hole excitations with respect to the Fermi sea incident in reservoir $\alpha$.~\footnote{The \textit{energy-resolved spectral current} should not be confused with the \textit{time-resolved current pulses} studied e.g. in Ref.~\onlinecite{Juergens11}.}  In the following, we focus on the zero-temperature regime. The Fermi functions are therefore replaced by sharp step functions, $[f(E_n)-f(E)]\rightarrow [\Theta(-E_n)-\Theta(-E)]$.

Finally, we are interested in the zero-frequency charge-current noise,~\cite{Blanter00} which is known to be sensitive to two-particle effects, 
\begin{eqnarray}\label{eq_defnoise}
\mathcal{P}_{\alpha\beta}&=&\frac{1}{2}\int_0^{\mathcal{T}} \frac{dt'}{\mathcal{T}} \int_{-\infty}^{\infty} d(t-t')\\ 
& &\big[\langle\hat{I}_{\alpha}(t)\hat{I}_{\beta}(t')+\hat{I}_{\beta}(t')\hat{I}_{\alpha}(t)\rangle-2\langle\hat{I}_{\alpha}(t)\rangle\langle\hat{I}_{\beta}(t')\rangle\big]\nonumber .
\end {eqnarray}
In the limit of zero temperature, the expression for the zero-frequency noise power assumes a rather compact form.
Substituting Eq.~(\ref{eq_defcurr}) into Eq.~(\ref{eq_defnoise}), we find
\begin{eqnarray}\label{eq_noise_zeroT}
&&\mathcal{P}_{\alpha\beta}=\\
&&\frac{e^2}{2h}\sum_{m=-\infty}^\infty\text{sign}(m)\int_{-m\hbar\Omega}^0dE\int_0^{\mathcal{T}} \frac{dt}{\mathcal{T}} \int_0^{\mathcal{T}} \frac{dt'}{\mathcal{T}} e^{im\Omega(t'-t)}\nonumber\\ 
&&\sum_{\gamma,\delta}\big[S^*_{\alpha\gamma}(t,E)S_{\alpha\delta}(t,E_m)S^*_{\beta\delta}(t',E_m)S_{\beta\gamma}(t',E)\big].\nonumber
\end{eqnarray}

In what follows all currents are evaluated at the detector situated at reservoir $\alpha=4$. We thus suppress the reservoir index, taking $i_4(E)\equiv i(E)$, $\bar{I}_4\equiv\bar{I}$, $\bar{J}_4\equiv\bar{J}$. Furthermore, we are interested in the cross-correlation function of charge currents, for which we have $\mathcal{P}_{34}=\mathcal{P}_{43}\equiv\mathcal{P}$. Note that the time average over one period will always include \textit{electron as well as hole} contributions from the different time-dependently driven SPSs. We will in the next sections separate the contributions by adding superscripts e and h to the considered quantities and by using  the variables $n_k^\text{e/h}$, previously introduced in the context of Eq.~(\ref{eq_adiabatic_S}), to highlight the origin of the different terms stemming from electron and hole contributions.

\section{Single-particle interference - wave packet picture}\label{sec_single}

It is instructive to first consider the situation, where SPS$_\text{B}$ is switched off and the signal injected into the MZI from SPS$_\text{A}$ leads to an interference pattern in the detected signal in reservoir $4$. The excess-energy distribution function (or spectral current) at the detector reads
\begin{subequations}
\begin{equation}\label{eq_i_MZI_A}
i_\text{MZI,A}(E,\Phi)=i^\text{cl}_\text{MZI,A}(E)+i^\text{int}_\text{MZI,A}(E,\Phi)
\end{equation}
where the classical part and the interference part, which oscillates as a function of the magnetic-flux dependent phase $\phi(E,\Phi)$, are given by
\begin{align}
i^\text{cl}_\text{MZI,A}(E)& =  (R_\text{L}R_\text{R}+T_\text{L}T_\text{R})\left[i^\text{e}_\text{A}(E)+i^\text{h}_\text{A}(E)\right]\label{eq_i_MZI_Acl}\\
i^\text{int}_\text{MZI,A}(E,\Phi)& =  - 2\gamma\cos\Delta\phi(E,\Phi)\left[i^\text{e}_\text{A}(E)+i^\text{h}_\text{A}(E)\right]\label{eq_i_MZI_Aint} .
\end{align}
\end{subequations}
Here, we have defined $\gamma=t^*_\text{L}r^{}_\text{L}t^{}_\text{R} r^*_\text{R}=\sqrt{T_\text{L}T_\text{R}R_\text{L}R_\text{R}}$. The excess-energy distribution function contains both electron- and hole-like contributions from the emission of the different types of particles from SPS$_\text{A}$. The particles injected by SPS$_\text{A}$ into the edge states are described by the excess-energy distribution functions~\cite{Keeling08}
\begin{eqnarray}
i^\text{e}_\text{A}(E) & = & \Theta(E)n^\text{e}_\text{A}2\Omega\sigma_\text{A} e^{- 2E \sigma_\text{A}/\hbar} \label{eq_i_SPSe}\\
i^\text{h}_\text{A}(E) & = &-\Theta(-E)n^\text{h}_\text{A}2\Omega\sigma_\text{A} e^{2E \sigma_\text{A}/\hbar}\label{eq_i_SPSh}\
\end{eqnarray}
of electron-like and hole-like excitations, with contributions in the positive, respectively the negative, energy range, only.  Note that, according to the definition given in Eq.~(\ref{eq_energy_distribution}), the excess-energy distribution function of the hole-like excitations, $i^\text{h}_{\alpha}(E)$, is always negative, which is consistent with the interpretation of a ``hole" as a missing electron in the Fermi sea.~\footnote{When introducing the magnetic field, which determines the direction of propagation of the chiral edge states, as an additional variable to the excess-energy distribution function, the equality $i^\text{e}_{\alpha}(E,\boldsymbol{B})=-i^\text{h}_{\alpha}(-E,-\boldsymbol{B})$ relates the excess-energy distribution function of electrons, $i^\text{e}_{\alpha}$, to the one of holes, $i^\text{h}_{\alpha}$.}
\begin{figure}[b]
\includegraphics[width=3.3in]{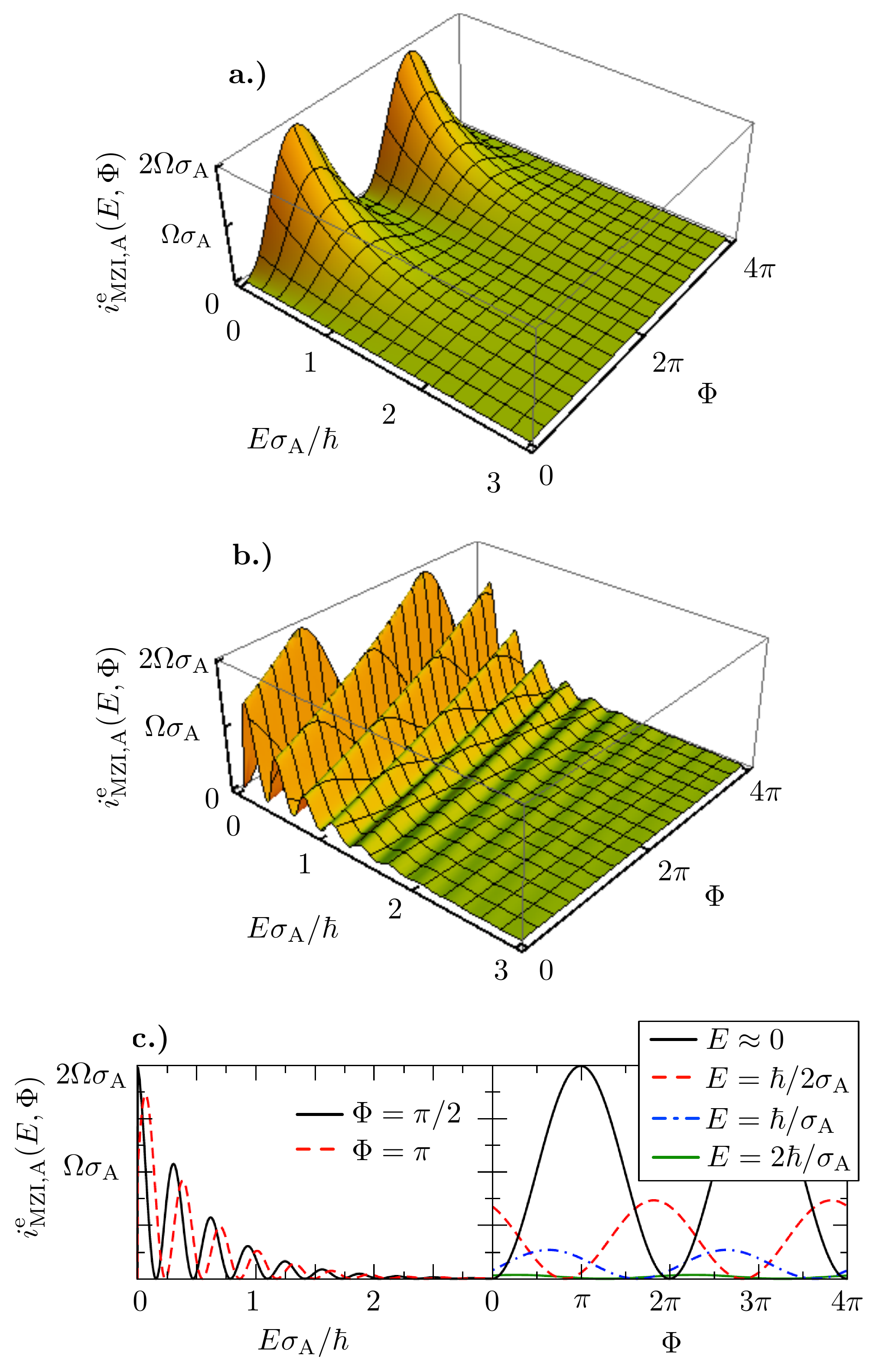}
\caption{Electronic part of the excess-energy distribution function, $i^\text{e}_\text{MZI,A}(E, \Phi)$ as a function of the energy $E$ in units of $\hbar/\sigma_\mathrm{A}$ and the magnetic-flux-dependent phase $\Phi$. a.)  Almost perfectly balanced interferometer, with $\Delta\tau=0.01\sigma_\text{A}$. b.) Unbalanced interferometer, with $\Delta\tau=20\sigma_\text{A}$.  c.) Cuts through the 3D plot of b.) at different energies,  $E$, and phases, $\Phi$. In all plots, the transmission probabilities are given by $T_\text{L}=T_\text{R}=0.5$.}
\label{fig_endis_MZI_A} 
\end{figure}

The term $i^\text{cl}_\text{MZI,A}(E)$, see Eq.~(\ref{eq_i_MZI_Acl}), is of classical nature and it is given by the sum of contributions from particles reaching the detector after travelling the upper or the lower arm with a probability $R_\text{L}R_\text{R}$, respectively $T_\text{L}T_\text{R}$.
In contrast, $i^\text{int}_\text{MZI,A}(E,\Phi)$, see Eq.~(\ref{eq_i_MZI_Aint}), shows the wave nature of the emitted signals.  It is due to the interference between waves propagating along the upper and the lower arms.  
In the almost perfectly balanced case, $\Delta\tau\leq\sigma_\text{A}$, shown in Fig.~\ref{fig_endis_MZI_A}~a.), we see the flux-dependence of the electronic contribution to the excess-energy distribution function,  $i^\text{e}_\text{MZI,A}(E,\Phi)$, which  is exponentially suppressed for increasing energies on the energy scale given by the inverse of the pulse width $\hbar/\sigma_\text{A}$.  In contrast, for a strongly  unbalanced interferometer,  $\Delta\tau\gg\sigma_\text{A}$, as shown in Fig.~\ref{fig_endis_MZI_A}~b.), also the energy-dependent part $E\Delta\tau/\hbar$ of the phase $\Delta\phi(E,\Phi)$ starts to play an important role leading to exponentially damped, fast energy-dependent oscillations in the spectral current. This goes along with a phase shift between the different energy contributions. In Fig.~\ref{fig_endis_MZI_A} c.), where we show phase- and energy-dependent cuts through the plot in Fig.~\ref{fig_endis_MZI_A}~b.), this behaviour is clearly visible.

\begin{figure}[t]
\includegraphics[width=3.3in]{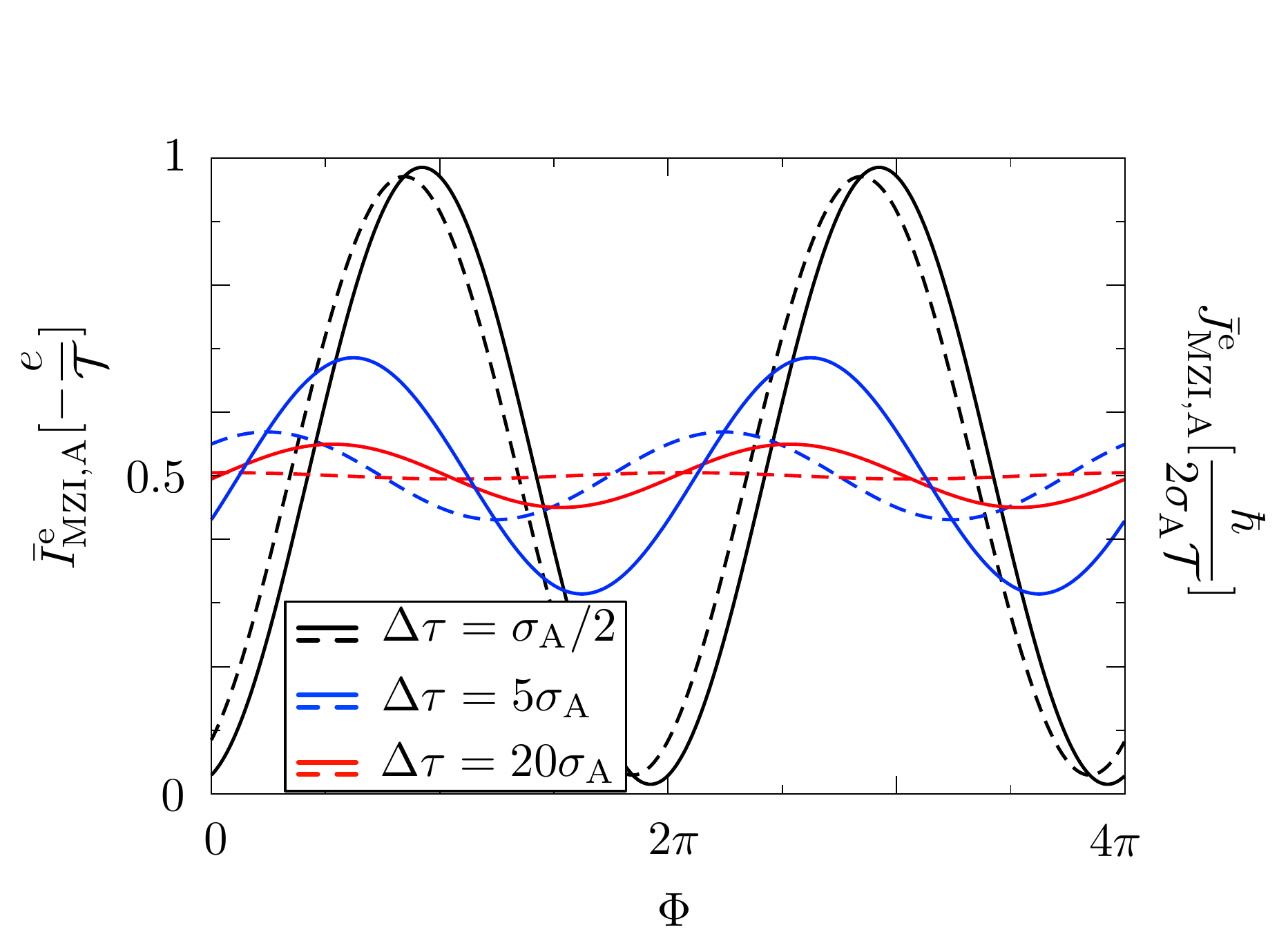}
\caption{Electronic part of the average charge current, $\bar{I}^\text{e}_\text{MZI,A}$, (full lines) and of the average energy current, $\bar{J}^\text{e}_\text{MZI,A}$, (dashed lines) as a function of the phase $\Phi$ for different values of the detuning $\Delta\tau$. The transmission probabilities are $T_\text{L}=T_\text{R}=0.5$.}
\label{fig_intcurrent_MZI_A} 
\end{figure}
The energy dependence of the interference part of the excess-energy distribution function is the electron analogue of  the so-called channelled spectrum known from optics.~\cite{Ferraro13} This energy dependence leads to dramatic differences for the charge and energy currents -- namely the energy-integrated quantities --  between the case of a balanced and a strongly unbalanced interferometer. 
The analytic results for the time-averaged charge and energy currents, consisting of the sum of an electronic and a hole-like contribution, are given by
\begin{eqnarray}\label{eq_curr_MZI}
&&\frac{\bar{I}_\text{MZI,A}}{-e/\mathcal{T}}   =  \left(R_\mathrm{L} R_\mathrm{R}+T_\mathrm{L} T_\mathrm{R}\right)\left(n_\mathrm{A}^\mathrm{e}-n_\mathrm{A}^\mathrm{h}\right)\label{eq_I_MZIA}\\
&& -2\gamma\mathbb{R}\mathrm{e}\left\{e^{-i\Phi}\left(n_\text{A}^\text{e}\frac{-2i\sigma_\mathrm{A}}{\Delta\tau-2i\sigma_\text{A}}-n_\text{A}^\text{h}\frac{2i\sigma_\mathrm{A}}{\Delta\tau+2i\sigma_\text{A}}\right)
\right\}\nonumber
\end{eqnarray}
\begin{eqnarray}
&& \frac{\bar{J}_\text{MZI,A}}{\hbar/(2\sigma_\text{A}\mathcal{T})}   =   \left(R_\mathrm{L} R_\mathrm{R}+T_\mathrm{L} T_\mathrm{R}\right)\left(n_\mathrm{A}^\mathrm{e}+n_\mathrm{A}^\mathrm{h}\right)
\label{eq_IE_MZIA} \\
&&-2\gamma\mathbb{R}\mathrm{e}
\left\{e^{-i\Phi}
\left(
n_\text{A}^\text{e}\left[\frac{-2i\sigma_\text{A}}{\Delta\tau-2i\sigma_\text{A}}\right]^2
\hspace{-0.2cm}+n_\text{A}^\text{h}\left[\frac{2i\sigma_\text{A}}{\Delta\tau+2i\sigma_\text{A}}\right]^2
\right)
\right\}
\nonumber.
\end{eqnarray}
These time-averaged charge and energy currents are obtained from the energy integral over the excess-energy distribution function. The equations show the sum of the electron and hole contributions, which are indicated by factors $n_\text{A}^\text{e}$ and $n_\text{A}^\text{h}$ stemming from different parts of the driving cycle. When considering a full period, both  $n_\text{A}^\text{e}$ and $n_\text{A}^\text{h}$ are equal to one. Fig.~\ref{fig_intcurrent_MZI_A} shows their electronic contributions only (a full 3D plot as function of $\Delta\tau$ and $\Phi$ is shown in Figs.~\ref{fig_charge_energy}~a.) and d.); equivalent results are found for the hole-like contributions). Importantly, when  $\Delta\tau\leq\sigma_\text{A}$, the interference pattern, observed in the excess-energy distribution function is clearly visible also in the charge and energy currents. However, when $\Delta\tau\gg\sigma_\text{A}$, the interference contributions to charge and energy currents are strongly suppressed. This suppression of the flux dependence can be understood as an averaging effect of the phase-shifted contributions  of the excess-energy distribution function at different energies. 

On the other hand, this suppression of interference is also a manifestation of the particle nature of the injected signal, made of a sequence of well-separated current pulses carrying exactly one electron or one hole. It has been shown in Refs.~\onlinecite{Haack11,Haack13} that the width in time of these current pulses,  $\sigma_\text{A}$,  is directly related to the \textit{single-particle coherence time} of electrons and holes.  The latter can be read out by measuring the visibility of the current signal detected at the output of an MZI: whenever the detuning of the interferometer, characterised by $\Delta\tau$, is much larger than the single-particle coherence time $\sigma_\text{A}$, the interference in the charge (and energy) current is suppressed. In this case the current pulses travelling along the upper arm and the lower arm arrive at the detector in well separated time  intervals and the signals from the two different paths are thus distinguishable. 

The coexistence of these two interpretations is consistent with the idea that, in quantum mechanics, a particle is described by a wave packet, composed of a superposition of plane waves at different energies.

Furthermore, from Eqs.~(\ref{eq_I_MZIA}) and (\ref{eq_IE_MZIA}), we see that the contributions for electrons and holes have different weights for finite detuning $\Delta\tau$. This is related to the different energies at which electron- and hole-like excitations occur and to the energy-filtering properties of the MZI. Consequently, as soon as the detuning is finite, the dc charge current at each of the two outputs is finite, even though the  charge current  injected by the SPS$_\text{A}$ into the MZI sums up to zero. As an additional result of the finite detuning, a phase shift with respect to the $\cos(\Phi)$-dependence is introduced.  The energy dependence of the excess-energy distribution function, namely the channelled spectrum, hence leads to charge and energy currents which are in general out of phase. 
Therefore, it is possible to tune the magnetic flux such that an electron is detected with a higher probability in reservoir 4, while the energy detected in reservoir 3 is on average larger than the one detected in reservoir 4 (and vice versa).
The different dependence of the phase shift in charge and energy currents as well as of the different suppression of the visibility as a function of the detuning can easily be seen by rewriting their interference contributions  as 
\begin{eqnarray}
&&\frac{\bar{I}^\text{e,int}_\text{MZI,A}}{-e/\mathcal{T}}=\label{eq_visi_I_MZIA}\\
&&-2\gamma\frac{2\sigma_\text{A}}{\sqrt{\Delta\tau^2+4\sigma_\text{A}^2}}\left(n^\text{e}_A\cos(\Phi+\psi^I) +n^\text{h}_A\cos(\Phi-\psi^I) \right)\nonumber
\end{eqnarray}
\begin{eqnarray}
&&\frac{\bar{J}^\text{e,int}_\text{MZI,A}}{\hbar/(2\sigma_\text{A}\mathcal{T})}=\label{eq_visi_IE_MZIA}\\
&& -2\gamma  \frac{4\sigma_\text{A}^2}{\Delta\tau^2+4\sigma^2}\left(n^\text{e}_A\cos(\Phi+\psi^J)+n^\text{h}_A\cos(\Phi-\psi^J)\right)\ .\nonumber
\end{eqnarray}
The different phase shifts are (where for the energy current we here give the explicit form for small detuning, $\Delta\tau<2\sigma_\text{A}$)
\begin{eqnarray}
\psi^I&=&\arctan\bigg(\frac{\Delta\tau}{2\sigma_\text{A}}\bigg)\\
\psi^J&=&\arctan\bigg(\frac{4\sigma_\text{A}\Delta\tau}{4\sigma_\text{A}^2-\Delta\tau^2}\bigg)\ .
\end{eqnarray}
Only when $\Delta\tau\rightarrow 0$, the phase difference $\Delta\phi$ becomes energy independent in Eq.~(\ref{eq_i_MZI_Aint}), and we find $\psi^I=\psi^J=0$. Consequently, charge and energy currents are then in phase. 

Since the energy current,  $\bar{J}=h^{-1}\int^{\infty}_{-\infty} dE  \ E\ i_{\alpha}(E)$, contains an additional factor  $E$ in the integrand with respect to the charge current, this quantity is more sensitive to the energy dependence of the distribution function. Thus, it is also more sensitive than the charge current to the variation of the interferometer imbalance showing interference suppression at smaller $\Delta\tau$ values, see Fig.~\ref{fig_intcurrent_MZI_A} for the electronic contributions to charge and energy currents. The visibility extracted from Eq.~(\ref{eq_visi_I_MZIA}) for the charge current in the case of symmetric transmission of the QPCs, namely $|I^{i,\text{int}}_\text{MZI,A}/I^{i,\text{cl}}_\text{MZI,A}|=2\sigma_\text{A}/\sqrt{\Delta\tau^2+4\sigma_\text{A}^2}$ indeed decays slower with $\Delta\tau$ than the visibility extracted from Eq.~(\ref{eq_visi_IE_MZIA}) for the energy current, namely $|J^{i,\text{int}}_\text{MZI,A}/J^{i,\text{cl}}_\text{MZI,A}|=4\sigma^2_\text{A}/(\Delta\tau^2+4\sigma_\text{A}^2)$.

An MZI fed by a non-adiabatically driven SPS has  recently been studied by Ferraro \textit{et al.}~\cite{Ferraro13} in the framework of Wigner functions. In that case the excess-energy distribution function of emitted particles, $i^\text{e/h}_\text{A}(E)$, is approximated by a Lorentzian function. The system shows a qualitatively similar behaviour to the one described here. A closely related work by Hofer and Flindt~\cite{Hofer14}  focuses on the propagation of multi-electron pulses through a Mach-Zehnder interferometer.

\section{Synchronised particle emission from two sources}\label{sec_synchronized}
We now come to the main subject of our work, the influence of the particle emission from SPS$_\text{B}$ on the interference pattern of the currents at the output of the MZI. It has been shown in Ref.~\onlinecite{Juergens11} that the interference pattern in the \textit{time-resolved current}, $\langle\hat{I}(t)\rangle$, detected at the output of the MZI is subject to a phase-shift, which can take values between $0$ and $2\pi$, depending on the emission time of electrons or holes from source B. This has as a consequence that the interference effects in the \textit{time-averaged current}, $\bar{I}$, detected at the output of the interferometer in every half period, get strongly suppressed when the emission of the particles is synchronised such that either particles emitted from SPS$_\text{A}$ can be absorbed at SPS$_\text{B}$ or that particles of the same kind can collide at QPC$_\text{R}$. This synchronisation of particles occurs as a perfect overlap of the time-resolved wave packets emitted from the two sources. A \textit{full absorption} thus can occur when $t_\text{A}^\text{e}+\tau_\text{d}/2=t_\text{B}^\text{h}$ (or $t_\text{A}^\text{h}+\tau_\text{d}/2=t_\text{B}^\text{e}$), which corresponds to $\Delta t_\text{d}^\text{eh}=0$ (or $\Delta t_\text{d}^\text{he}=0$), together with $\sigma_\text{A}=\sigma_\text{B}$. A \textit{full collision} of electrons (or holes) can occur when $t_\text{A}^\text{e}+\tau_\text{u}=t_\text{B}^\text{e}+\tau_\text{d}/2$ (or $t_\text{A}^\text{h}+\tau_\text{u}=t_\text{B}^\text{h}+\tau_\text{d}/2$), which corresponds to $\Delta t_\text{u}^\text{ee}=0$ (or $\Delta t_\text{u}^\text{hh}=0$), together with $\sigma_\text{A}=\sigma_\text{B}$. 

Interestingly, the conditions for the averaging of the time-resolved currents, leading to a full suppression of the interference effects in the detected charge, allow for a particularly interesting interpretation, which has been put forward in Ref.~\onlinecite{Juergens11}. This interpretation is based on which-path information which can be acquired in the case that particle collisions or absorptions occur due to an appropriate synchronization of the two SPSs. In order to introduce this interpretation in a nutshell, let us for the moment assume for simplicity that the QPCs defining the MZI are both semi-transparent.
   
We first consider the situation where SPS$_\text{A}$ emits an electron and SPS$_\text{B}$ a hole. Whenever the condition $\Delta t_\text{d}^\text{eh}=0$ is fulfilled, no particle arrives at any of the outputs, when the electron emitted from source A takes the lower arm of the MZI and gets absorbed. When the particle emitted from A takes the upper arm, the average charge remains to be equal to zero, however fluctuations occur.  This leads to which-path information suppressing the interference effect: whenever an electron or a hole is detected in one of the detectors, we can conclude that the electron emitted from SPS$_\text{A}$ took the upper arm.
 
\begin{figure}[t]
\includegraphics[width=3.3in]{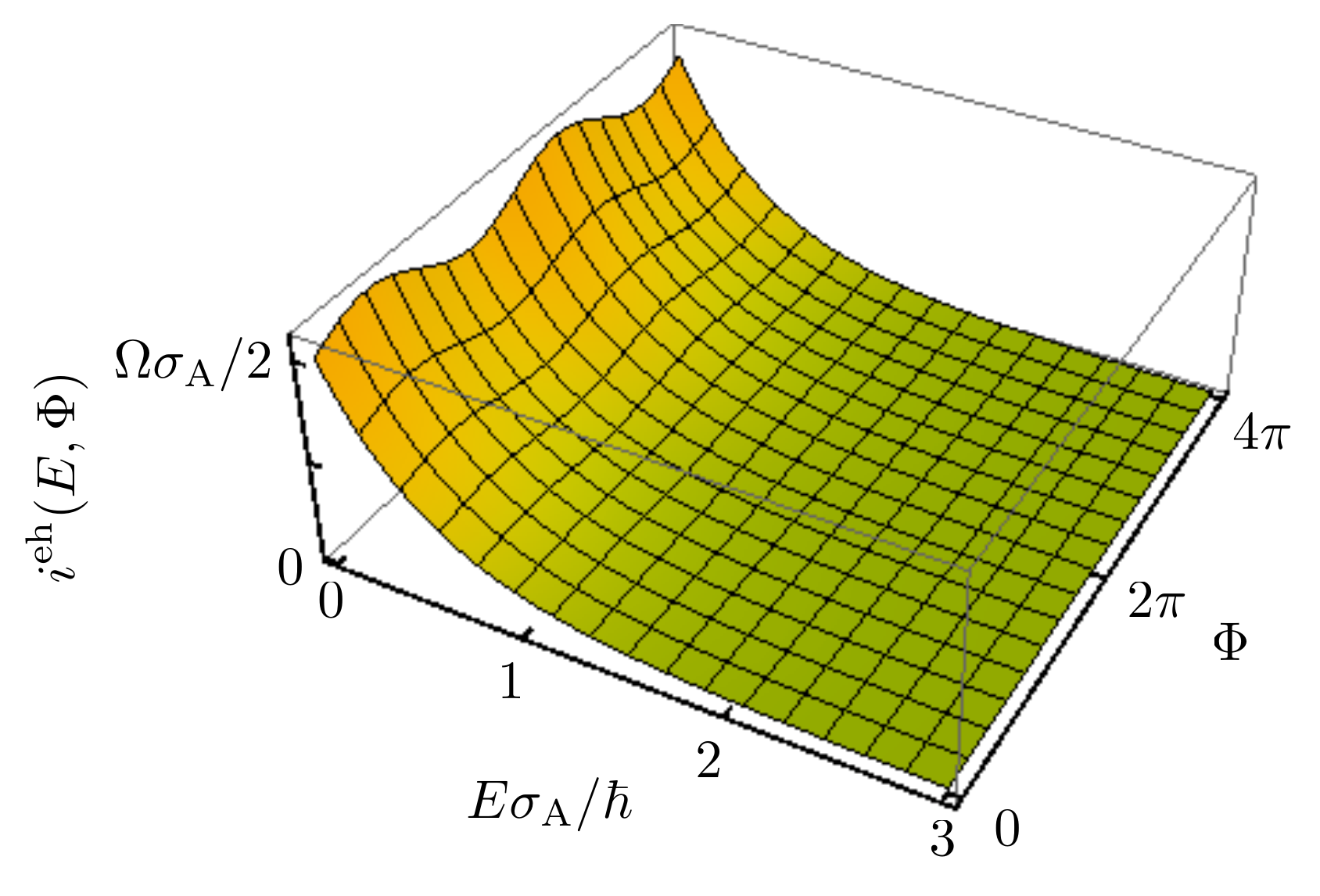}
\caption{Energy-distribution function, $i^\text{eh}(E, \Phi)$, shown for positive values of the energy $E$ only, in the regime where absorptions of electrons emitted by A are possible through the emission of holes from B depending on the time difference $\Delta t^\text{eh}_\text{d}$. Here we take $\Delta t^\text{eh}_\text{d}=0.1\sigma_\text{A}$ and show $i^\text{eh}(E, \Phi)$ as a function of the energy $E$ in units of $\hbar/\sigma_\mathrm{A}$ and the magnetic-flux-dependent phase $\Phi$. The interferometer is almost perfectly balanced, $\Delta\tau=0.01\sigma_\text{A}$, the pulse widths are assumed to be equal, $\sigma_\text{A}=\sigma_\text{B}$, and the transmission probabilities are given by $T_\text{L}=T_\text{R}=0.5$.}
\label{fig_iabs} 
\end{figure}

Equally, when both SPSs emit electrons and the condition  $\Delta t_\text{u}^\text{ee}=0$ is fulfilled, these two electrons could collide at QPC$_\text{R}$. When the electron emitted from source A travels along the upper arm of the MZI, the two electrons - being in the same state - would have to be scattered to the two opposite outputs of the MZI at QPC R, due to fermion statistics;~\cite{Olkhovskaya08} in the case that the particle emitted from A takes the lower arm of the MZI both particles can go to both outputs randomly. This means that the average charge in each detector is always $-e$ independently of the traversed path, however only when the electron from SPS$_\text{A}$ took the lower arm, fluctuations can occur. This again leads to which-path information leading to an interference suppression: whenever 0 or 2 electrons arrive in one of the detectors, we can conclude that the electron from SPS$_\text{A}$ took the lower arm. 

Note that this setup is very different from MZIs where an interference suppression is reached by placing a voltage probe~\cite{Buttiker88} in one of the interferometer arms.~\cite{Marquardt04PRL,Marquardt04PRB,Roulleau09} A voltage probe acts as a which-path detector itself and leads to dephasing. However, the presence of SPS$_\text{B}$ leads to a coherent suppression of interference and which-path information can be acquired only at the detectors at the outputs of the MZI, thanks to the synchronized emission of particles from SPS$_\text{B}$. 

In the following, in addition to the charge current we will investigate also the spectral current and the energy current of the emitted signals as well as the charge-current noise with the aim to extend the understanding of the impact of the above described multi-particle effects on the MZI signal.

\subsection{Spectral properties}\label{sec_2_spectral}

We start by considering the spectral currents 
 for the case where one source emits an electron and one source emits a hole, allowing for the absorption of particles at SPS$_\text{B}$,
 as well as the case where  
 both sources, SPS$_\text{A}$ and SPS$_\text{B}$, emit the same kind of particles, allowing for possible collisions between particles  in one half period.
 The synchronised emission from the two sources goes along with inelastic scattering processes. More specifically, scattering at the time-dependently driven SPS$_\text{B}$ results in an energy increase or decrease in the scattering process. This  leads to  a deformation of the spectral distribution of the current as will be shown in the following.

\subsubsection{Absorption  of particles} 

In the case where particles of opposite type emitted from the two sources are detected in the same half period, absorptions can occur at source B and the spectral current is given by
\begin{align}\label{eq_spec_abs}
&i^\text{eh}(E,\Phi)  =  
 R_\mathrm{L}R_\mathrm{R}i_\mathrm{A}^\mathrm{e}(E)+
		R_\mathrm{L}T_\mathrm{R}i_\mathrm{B}^\mathrm{h}(E)\\
&+T_\mathrm{L}T_\mathrm{R}	\left(
i_\mathrm{B}^\mathrm{h}(E)+i_\mathrm{A}^\mathrm{e}(E)
	\right)\left(1-
		\frac{
			4\sigma_\mathrm{A}\sigma_\mathrm{B}}
			{{\Delta t_\mathrm{d}^\mathrm{eh}}^2+\left(\sigma_\mathrm{A}+\sigma_\mathrm{B}\right)^2}\right)
\nonumber\\
& -2\gamma i_\mathrm{A}^\mathrm{e}(E)\mathbb{R}\mathrm{e}\left\{
e^{-i\Phi}e^{-iE\Delta\tau/\hbar}
\left(
		1-\frac{
			2i\sigma_\mathrm{B}}
			{{\Delta t_\mathrm{d}^\mathrm{eh}}+i\left(\sigma_\mathrm{A}+\sigma_\mathrm{B}\right)}
\right)
\right\}
 .\nonumber
\end{align}
From now on, for observables calculated for the MZI with two sources, we drop the subscript indicating the presence of the MZI and the number of working sources, the latter being evident from the superscript $ij$ for the type of particle $i=\text{e,h}$ emitted from SPS$_\text{A}$ and the type of particle $j=\text{e,h}$ emitted from SPS$_\text{B}$. Here, we show the case where SPS$_\text{A}$ emits an electron and SPS$_\text{B}$ a hole ($n_\text{A}^\text{e}=n_\text{B}^\text{h}=1$ and $n_\text{A}^\text{h}=n_\text{B}^\text{e}=0$); the opposite case is shown in Appendix~\ref{app_analytic_spec}.

Far away from the condition, $\Delta t^\text{eh}_\text{d}=0$ and $\sigma_\text{A}=\sigma_\text{B}$, the two particles are emitted independently, such that the electron emitted from SPS$_\text{A}$ is not in the vicinity of SPS$_\text{B}$, when a hole emission occurs at the latter. Then the expression given in Eq.~(\ref{eq_spec_abs}) reduces to the sum of the separate contributions of the two sources, namely for the hole emitted from SPS$_\text{B}$ and transmitted at QPC$_\text{R}$, $T_\text{R}i^\text{h}_\text{B}(E)$, and the electron term containing interference effects, given in Eq.~(\ref{eq_i_MZI_A}).

The collision of an electron emitted from SPS$_\text{A}$ and a hole emitted from SPS$_\text{B}$ at the position of the latter source (which is equivalent to the absorption of electrons emitted from SPS$_\text{A}$ at SPS$_\text{B}$) can occur when the time difference $\Delta t_\text{d}^\text{eh}$ is of the order of the width of the associated time-resolved current pulses $\sigma_\text{A},\sigma_\text{B}$. It leads to a cancellation of the contribution of the current travelling along the lower arm in an \textit{energy-independent} manner, depending only on how accurately the absorption conditions, $\Delta t_\text{d}^\text{eh}= 0$ and $\sigma_\text{A}=\sigma_\text{B}$, are fulfilled. 
Equally, the suppression of the interference part of the current takes place in a way which is independent of the energy $E$.
It becomes evident also from Fig.~\ref{fig_iabs}, where the electronic part of this spectral current is shown as a function of energy and of the magnetic-flux dependent phase. Indeed, the amplitude of the flux-dependent oscillations is suppressed with respect to the case where $\Delta t_\text{d}^\text{eh}\gg\sigma_\text{A/B}$ -- the latter being equivalent to the case of an emission from A only, while source B is switched off, see Fig.~\ref{fig_endis_MZI_A}~a.).

\subsubsection{Collision of particles of the same kind} 

 In the case where particles of the same type emitted from both sources are detected in one half period, we find for the spectral current
\begin{widetext}
\begin{eqnarray}\label{eq_spec_coll}
&& i^\mathrm{ee}(E,\Phi)  =  R_\mathrm{L}R_\mathrm{R}i_\mathrm{A}^\mathrm{e}(E)+T_\mathrm{L}T_\mathrm{R}i_\mathrm{A}^\mathrm{e}(E)
\mathbb{R}\mathrm{e}\left\{
		1+\frac{4\sigma_\mathrm{A}\sigma_\mathrm{B}}{{\Delta t_\mathrm{d}^\mathrm{ee}}^2+(\sigma_\mathrm{A}-\sigma_\mathrm{B})^2}
-2i\sigma_\mathrm{B}\frac{\Delta t_\mathrm{d}^\mathrm{ee}-i(\sigma_\mathrm{A}+\sigma_\mathrm{B})}{{\Delta t_\mathrm{d}^\mathrm{ee}}^2+(\sigma_\mathrm{A}-\sigma_\mathrm{B})^2}e^{-iE\left(\Delta t_\mathrm{d}^\mathrm{ee}+i\left(\sigma_\mathrm{A}-\sigma_\mathrm{B}\right)\right)/\hbar}\right\}
\nonumber\\
&&+R_\mathrm{L}T_\mathrm{R}i_\mathrm{B}^\mathrm{e}(E) +T_\mathrm{L}T_\mathrm{R}i_\mathrm{B}^\mathrm{e}(E)
\mathbb{R}\mathrm{e}\left\{
		1+\frac{4\sigma_\mathrm{A}\sigma_\mathrm{B}}{{\Delta t_\mathrm{d}^\mathrm{ee}}^2+(\sigma_\mathrm{A}-\sigma_\mathrm{B})^2}
-2i\sigma_\mathrm{A}\frac{\Delta t_\mathrm{d}^\mathrm{ee}-i(\sigma_\mathrm{A}+\sigma_\mathrm{B})}{{\Delta t_\mathrm{d}^\mathrm{ee}}^2+(\sigma_\mathrm{A}-\sigma_\mathrm{B})^2}e^{-iE\left(\Delta t_\mathrm{d}^\mathrm{ee}+i\left(\sigma_\mathrm{B}-\sigma_\mathrm{A}\right)\right)/\hbar}\right\}\nonumber\\
&&-2\gamma i_\mathrm{A}^\mathrm{e}(E)\mathbb{R}\mathrm{e}\left\{
e^{-i\Phi}e^{-iE\Delta\tau}
\left[1+\frac{2i\sigma_\mathrm{B}}{\Delta t_\mathrm{d}^\mathrm{ee}+i\left(\sigma_\mathrm{A}-\sigma_\mathrm{B}\right)}
\left(1-e^{-iE\left(\Delta t_\mathrm{d}^\mathrm{ee}+i\left(\sigma_\mathrm{A}-\sigma_\mathrm{B}\right)\right)/\hbar}\right)
\right]
\right\}\ ,
\end{eqnarray}
\end{widetext}
where we here show the electron part, only; the hole contribution is given in Appendix~\ref{app_analytic_spec}.

\begin{figure}[bt]
\includegraphics[width=3.3in]{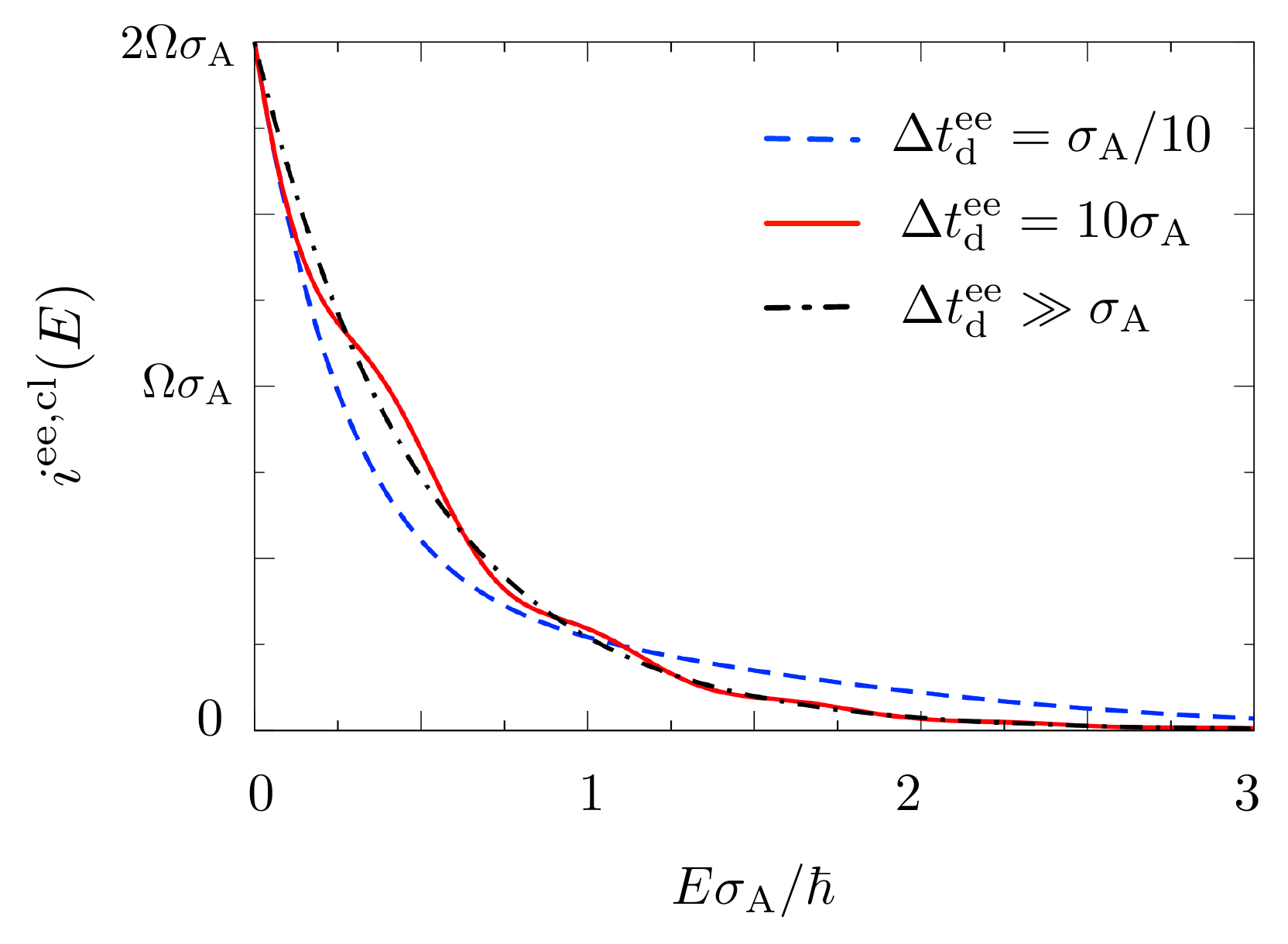}
\caption{Classical part of the excess-energy distribution function, $i^\text{ee,cl}(E)$, in the regime where collisions between particles of the same kind are possible depending on the time difference $\Delta t^\text{ee}_\text{d}$. We show the electronic contribution as a function of the energy $E$ in units of $\hbar/\sigma_\mathrm{A}$. We take $\sigma_\text{A}=\sigma_\text{B}$  and the transmission probabilities are given by $T_\text{L}=T_\text{R}=0.5$.}
\label{fig_icoll_cl} 
\end{figure}

The classical part, $i^\mathrm{ee,cl}(E)$, is given by the expression in the first two lines of Eq.~(\ref{eq_spec_coll}). Again, it reduces to the sum of the  single-particle contributions, namely the sum of $T_\mathrm{R}i_\mathrm{B}^\mathrm{e}$ and of the expression in Eq.~(\ref{eq_i_MZI_Acl}), when $\Delta t^\mathrm{ee}_\text{d}\gg\sigma_\mathrm{A},\sigma_\mathrm{B}$. 
The resulting exponential behaviour of the spectral current is represented by the black (dashed-dotted line) in Fig.~\ref{fig_icoll_cl}.
However, if the tuning of the emission times from SPS$_\text{A}$ and SPS$_\text{B}$ is such that particles could collide at SPS$_\text{B}$, in other words, if there is an overlap of the \textit{time-resolved current pulses} emitted from the two sources and the difference of the emission times, $\Delta t^\mathrm{ee}_\text{d}$, is of the order of the width of the current pulses, then energy-dependent oscillations occur in the classical part of the \textit{spectral current} on a scale given by the inverse of the  time difference, $\hbar/\Delta t^\mathrm{ee}_\text{d}$.  This oscillation on top of the energy-dependent exponential decay of the spectral current is a result of the complex exponential factor in the last term of the first two lines of Eq.~(\ref{eq_spec_coll}). Importantly, its amplitude gets suppressed for large time differences. Therefore the \textit{amplitude} of the oscillations is largest close to the collision condition $\Delta t^\mathrm{ee}_\text{d}=0$, while the \textit{frequency} of the oscillations is reduced. 
This behaviour becomes apparent from the red (full) line in the plot shown in Fig.~\ref{fig_icoll_cl} where damped oscillations are visible. The oscillations of the blue (dashed) line are hardly visible due to the small oscillation frequency. It is this complex energy dependence at the scale $\hbar/\Delta t_\text{d}^\text{ee}$, which leads to the fact that the classical part of the energy-integrated, average charge current is insensitive to collisions of particles at SPS$_\text{B}$, while an increase of the classical part of the energy current is observed when two  particles are emitted on top of each other at SPS$_\text{B}$.~\cite{Moskalets09}

This behaviour is very different from the energy-independent suppression of parts of the spectral current in the regime of possible particle absorptions.

\begin{figure}[bt]
\includegraphics[width=3.3in]{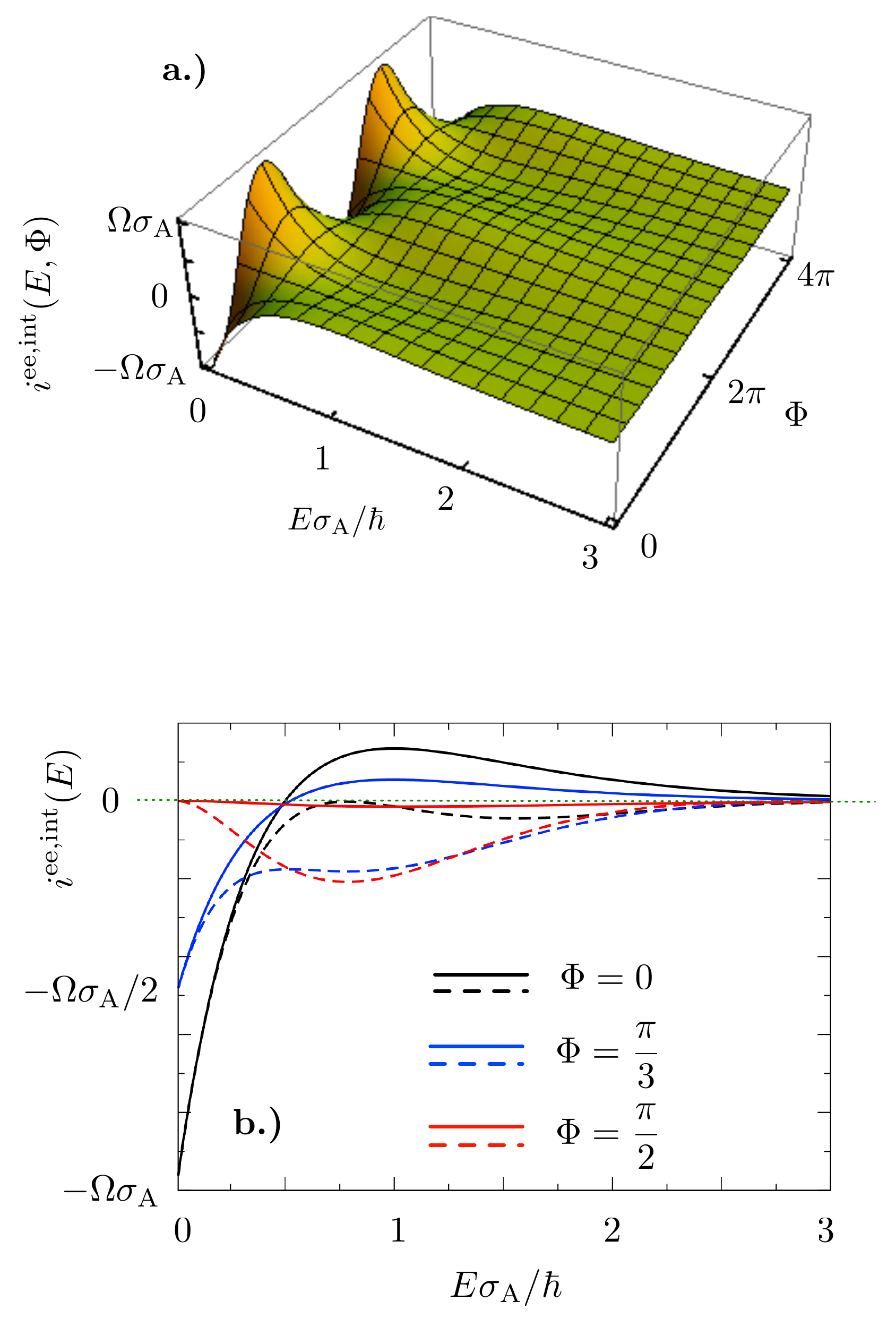}
\caption{Interference part of the excess-energy distribution function, $i^\text{ee,int}(E, \Phi)$, in the regime where collisions between particles of the same kind are possible depending on the time difference $\Delta t^\text{ee}_\text{d}$. We show the electron contribution a.) as a function of the energy $E$ in units of $\hbar/\sigma_\mathrm{A}$ and the magnetic-flux-dependent phase $\Phi$ close to collision $\Delta t^\text{ee}_\text{d}=0.1\sigma_\text{A}$ and b.)  as a function of the energy $E$ for three different flux values and for  $\Delta t^\text{ee}_\text{d}=0.1\sigma_\text{A}$ (full lines) and  $\Delta t^\text{ee}_\text{d}=2\sigma_\text{A}$ (dashed lines). The interferometer is almost perfectly balanced, $\Delta\tau=0.01\sigma_\text{A}$, the pulse widths are assumed to be equal, $\sigma_\text{A}=\sigma_\text{B}$, and the transmission probabilities are given by $T_\text{L}=T_\text{R}=0.5$.}
\label{fig_icoll_int} 
\end{figure}

The interference contribution, $i^\mathrm{ee,int}(E)$, is given by the third line of Eq.~(\ref{eq_spec_coll}) and it is shown in Fig.~\ref{fig_icoll_int}. Far from the collision condition, this contribution stems from the signal emitted from \textit{source} A \textit{only}, where it equals Eq.~(\ref{eq_i_MZI_Aint}). When the particles from SPS$_\text{A}$ and SPS$_\text{B}$ are emitted such that collisions between them are possible at SPS$_\text{B}$, oscillations with \textit{two competing time-scales} appear, namely the time-scale of the collision condition, $\Delta t^\mathrm{ee}_\text{d}$, and the time-scale related to the detuning of the interferometer, $\Delta\tau$. Again, oscillations on the energy scale given by $\hbar/\Delta t^\mathrm{ee}_\text{d}$ are suppressed for large time differences $\Delta t^\mathrm{ee}_\text{d}$. Note once more, that this is however very different from the absorption case where the time-scale of the absorption condition enters in a fully energy-independent manner. For an almost perfectly balanced interferometer, $\Delta\tau\ll\sigma_\text{A}$, the interference contribution to the spectral current is shown as a function of the energy and the flux-dependent phase in Fig.~\ref{fig_icoll_int}~a.), exhibiting slow oscillations on the scale $\hbar/\Delta t^\mathrm{ee}_\text{d}$, where we here chose the case close to the collision condition, $\Delta t^\mathrm{ee}_\text{d} =0.1\sigma_\text{A}$. In Fig.~\ref{fig_icoll_int}~b.) cuts through the three-dimensional plot of Fig.~\ref{fig_icoll_int}~a.) are shown as a function of energy for different values of the phase, $\Phi$. We compare these curves with the case slightly farther away from the collision condition, where the modulation on the energy scale given by $\hbar/\Delta t^\mathrm{ee}_\text{d}$ becomes more obvious. Interestingly, the areas enclosed by the curves below and above the energy-axis (indicated by the green dotted line in Fig.~\ref{fig_icoll_int}~b.)) close to the collision condition, $\Delta t^\mathrm{ee}_\text{d} =0.1\sigma_\text{A}$, sum up to a value close to zero independently of the value of the magnetic flux entering the phase $\Phi$. We will see in the following section, Section~\ref{sec_current}, that this leads to a suppression of the interference in the (energy-integrated) charge current, when the two sources are adequately synchronised. However, as soon as the time difference $\Delta t^\mathrm{ee}_\text{d}$ is increased while keeping the interferometer balanced, $\Delta\tau\approx0$, the sum of the enclosed areas becomes flux dependent, as can be seen from the dashed lines in Fig.~\ref{fig_icoll_int}~b.).

\subsection{Charge current}\label{sec_current}

\begin{figure*}
\centering
\includegraphics[width=\textwidth]{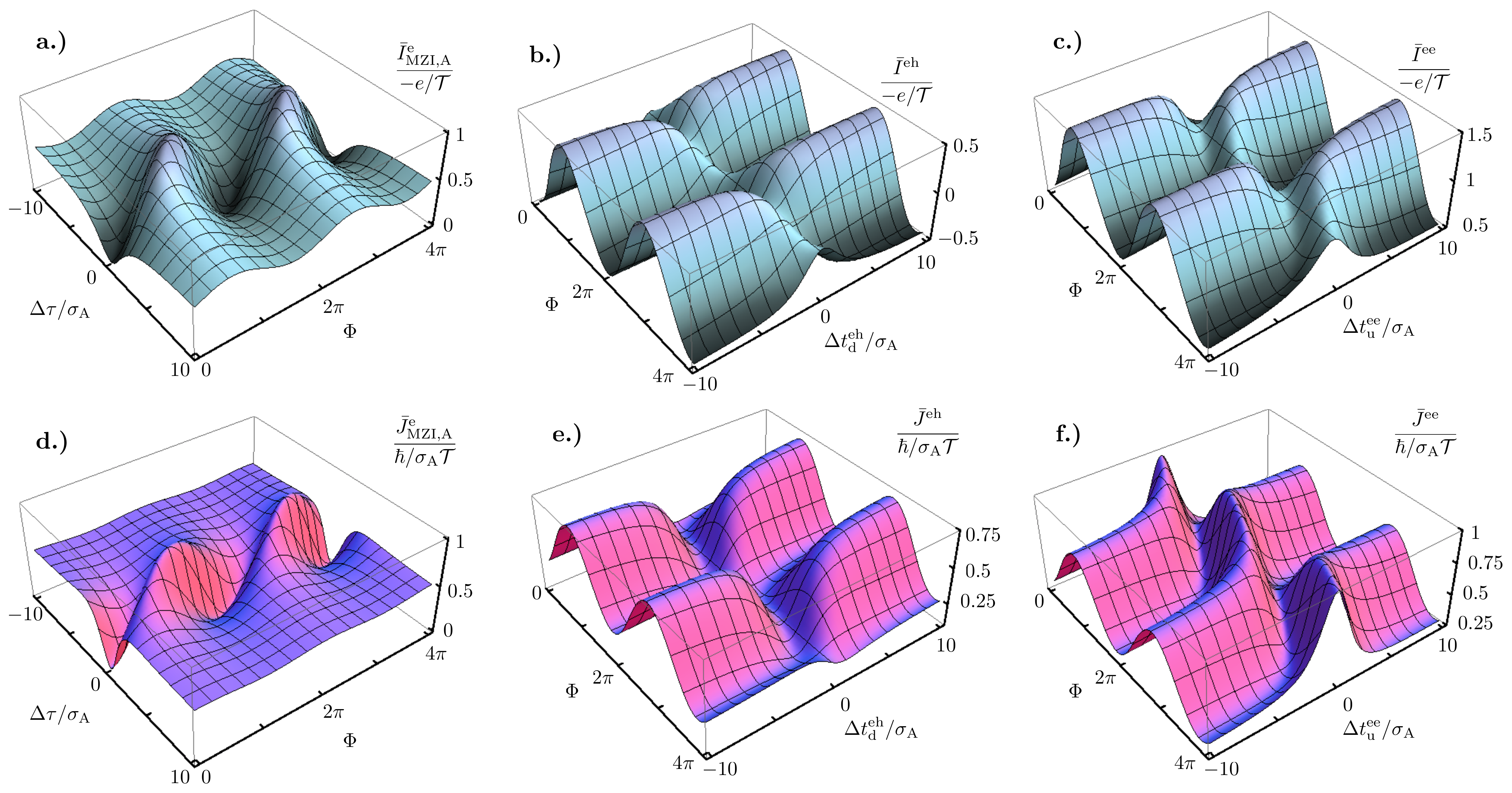}
\caption{Charge and energy current detected at reservoir~4 (output of the interferometer), with symmetric transmission of QPC$_\text{L}$ and QPC$_\text{R}$. a.) Charge current of an MZI fed by SPS$_\text{A}$ only, as a function of the MZI detuning $\Delta\tau$ in units of the current-pulse width $\sigma_\text{A}$ and as a function of the magnetic flux $\Phi$. b.) Charge current of a slightly detuned MZI, $\Delta\tau=0.5\sigma_\text{A}$  fed by an electron from SPS$_\text{A}$ and a hole from SPS$_\text{B}$ as a function of the time difference $\Delta t_\text{d}^\text{eh}$ in units of the pulse width $\sigma_\text{A}=\sigma_\text{B}$ and the magnetic flux-dependent phase $\Phi$. c.) Charge current of a slightly detuned MZI, $\Delta\tau=0.5\sigma_\text{A}$  fed by an electron both  from SPS$_\text{A}$ and  SPS$_\text{B}$ as a function of the time difference $\Delta t_\text{u}^\text{ee}$ in units of the pulse width $\sigma_\text{A}=\sigma_\text{B}$ and the magnetic flux-dependent phase $\Phi$.  d.)-f.) Energy currents for the same situations shown in a.)-c.).
}
\label{fig_charge_energy} 
\end{figure*}

The energy-dependent interference occurring in the previously studied spectral currents is equivalent to what is known as a \textit{channelled spectrum} from optics. The behaviour of the charge end energy currents, which are given by the energy averages of the spectral currents multiplied by the charge, respectively the energy, see Eqs.~(\ref{eq_current_density}) and (\ref{eq_energy_density}), can therefore be understood based on the previous investigations. Here, we start with the presentation of the charge current which is found in one half period in which an electron emitted from SPS$_\text{A}$ and a hole emitted from SPS$_\text{B}$ are detected in reservoir~4 (namely taking $n^\text{e}_\text{A}=n^\text{h}_\text{B}=1$ and $n^\text{h}_\text{A}=n^\text{e}_\text{B}=0$), allowing for the absorption of particles if $\Delta t_\text{d}^\text{eh}\approx 0$. The charge current is then given by
\begin{eqnarray}\label{eq_charge_abs}
&&\frac{\bar{I}^\text{eh}}{-e/\mathcal{T}}  = R_\text{L}R_\text{R}+T_\text{L}T_\text{R}-T_\text{R}\\
&& -2\gamma\mathbb{R}\mathrm{e}\left\{
e^{-i\Phi}\frac{-2i\sigma_\text{A}}{\Delta\tau-2i\sigma_\text{A}}
\left(1-\frac{2i\sigma_\text{B}}{\Delta t_\text{d}^\text{eh}+i\left(\sigma_\text{A}+\sigma_\text{B}\right)}\right)
\right\}\nonumber\ .
\end{eqnarray}
We find that only the interference part of the charge current is affected by the synchronisation of the particle emission from the two sources. The dependence of the spectral current on the time-difference $\Delta t^\text{eh}_\text{d}$, see Eq.~(\ref{eq_spec_abs}), thus cancels out in the classical part.
The factor leading to the maximum of interference for a balanced MZI, $\Delta\tau\rightarrow0$, in the absence of absorptions, and the factor suppressing the interference in case of absorptions, $\Delta t_\text{d}^\text{eh}$, are of very similar nature, both leading to a Lorentzian-type structure together with a phase shift at the maximum/minimum of their contribution. This similarity becomes also obvious when comparing Figs.~\ref{fig_charge_energy}~a.) and b.) which bring out the two effects.

The insensitivity of the classical part of the current to absorptions as well as the suppression of interference can  on one hand be interpreted as the result of an energy average of the spectral current given in Eq.~(\ref{eq_spec_abs}). A physically more insightful interpretation can however be given based on a particle picture of the injected signals. 
As explained in more detail in the beginning of this section, see also Ref.~\onlinecite{Juergens11}, the \textit{average} charge current of each classical path is not affected by an absorption - in other words, an electron and a hole carry in total no charge independently of whether they recombine in an absorption process or not. However,  the absorption of an electron by an emitted hole suppresses the fluctuations in the charge current. This difference in fluctuations depending on the arm that the particle emitted from SPS$_\text{A}$ took yields which-path information leading to an interference suppression.  This suppression of fluctuations in the case of absorptions is shown in a detailed study of the noise in Sec.~\ref{sec_noise}.

The charge current detected in the half period in which holes emitted from SPS$_\text{A}$ can be absorbed at SPS$_\text{B}$ behaves similarly as the one for the opposite case and it is given by
\begin{eqnarray}
&&\frac{\bar{I}^\text{he}}{-e/\mathcal{T}}  = -R_\text{L}R_\text{R}-T_\text{L}T_\text{R}+T_\text{R}\\
&& +2\gamma\mathbb{R}\mathrm{e}\left\{
e^{-i\Phi}\frac{2i\sigma_\text{A}}{\Delta\tau+2i\sigma_\text{A}}
\left(1-\frac{-2i\sigma_\text{B}}{\Delta t_\text{d}^\text{he}-i\left(\sigma_\text{A}+\sigma_\text{B}\right)}\right)
\right\}\nonumber .
\end{eqnarray}
The difference with respect to Eq.~(\ref{eq_charge_abs}), is given by a sign difference due to the contribution of oppositely charged particles and by a different phase which enters both through the factor stemming from the detuning properties of the MZI as well as from the factor describing the synchronisation of particles. As a consequence from this phase shift between hole and electron contribution, the charge current detected at reservoir~4 during a whole period, does not vanish (even though the total current injected into the MZI is zero) and it is given by
\begin{eqnarray}
&&\frac{\bar{I}^\text{eh+he}}{-e/\mathcal{T}}  = 2\gamma
\sin(\Phi)\frac{4\sigma_\text{A}\Delta\tau}{\Delta\tau^2+4\sigma_\text{A}^2}
\frac{\Delta t_\text{d}^2+\sigma_\text{A}^2-\sigma_\text{B}^2}{\Delta t_\text{d}^2+\left(\sigma_\text{A}+\sigma_\text{B}\right)^2}\nonumber\\
&&+2\gamma
\sin(\Phi)\frac{4\sigma_\text{A}^2}{\Delta\tau^2+4\sigma_\text{A}^2}
\frac{4\sigma_\text{B}\Delta t_\text{d}}{\Delta t_\text{d}^2+\left(\sigma_\text{A}+\sigma_\text{B}\right)^2}\ .
\end{eqnarray}
Here we assume that $\Delta t_\text{d}^\text{eh}=\Delta t_\text{d}^\text{he}\equiv\Delta t_\text{d}$ for simplicity. 
Both contributing terms depend on the energy-filtering properties of the MZI due to finite detuning $\Delta\tau$ and on the synchronised emission of multiple particles leading to a modification of the channelled spectrum of the device. The first of these terms is finite only for finite detuning, the other one occurs only when the emission of the two particles of opposite type is slightly detuned, $\Delta t_\text{d}\neq0$. Interestingly, the latter term is finite also when the detuning is zero: in this case the total charge current at the output of the MZI both due to SPS$_\text{A}$ alone and due to SPS$_\text{B}$ alone would vanish. However, when the sources are synchronised  such that $\Delta t_\text{d}\approx\sigma_\text{A},\sigma_\text{B}$, the term survives showing features due to two-particle effects in the dc charge current.

We now consider the case where an electron from each of the two SPSs arrives at the detector in the same half period. The average charge current in this case is
\begin{eqnarray}\label{eq_charge_coll}
&&\frac{\bar{I}^\text{ee}}{-e/\mathcal{T}}  = R_\text{L}R_\text{R}+T_\text{L}T_\text{R}+T_\text{R}\\
&& -2\gamma\mathbb{R}\mathrm{e}\left\{
e^{-i\Phi}\frac{-2i\sigma_\text{A}}{\Delta\tau-2i\sigma_\text{A}}
\left(1-\frac{-2i\sigma_\text{B}}{\Delta t_\text{u}^\text{ee}-i\left(\sigma_\text{A}+\sigma_\text{B}\right)}\right)
\right\}\nonumber\ .
\end{eqnarray}
Also in the expression for $\bar{I}^\text{ee}$, the classical contribution is independent of the synchronisation of the two sources; in contrast, the  interference part of the time-averaged charge current is sensitive to the collision of particles at QPC$_\text{R}$. This can again be understood as an energy-average of the synchronisation-dependent spectral current. Note however, that while the structure of the expression given in Eq.~(\ref{eq_charge_coll}) is very similar to the one for the absorption case, the corresponding spectral currents have very different behaviours. In particular, the fact that the time-scale $\Delta t_\text{d}^\text{ee}$ (for the emission of an electron at SPS$_\text{B}$ on top of the one from SPS$_\text{A}$) introduces an energy-dependent oscillation into the spectral current is important here: together with the energy-dependent oscillation governed by the time-scale of the detuning $\Delta\tau$ it leads to  features at the collision condition $\Delta t_\text{u}^\text{ee}\approx 0$, when the energy integration of the spectral current is performed to obtain the average charge current.

The interpretation of these facts is again more intuitive when resorting to an explanation based on a particle picture.
When the electron emitted from SPS$_\text{A}$ travels along the upper arm and the collision condition,  $\Delta t_\text{u}^\text{ee}$ and $\sigma_\text{A}=\sigma_\text{B}$, is fulfilled, it collides with the electron emitted from SPS$_\text{B}$ leading to the transmission of exactly one electron to each MZI output. When the electron  emitted from SPS$_\text{A}$ takes the lower arm, the charge in the two MZI outputs fluctuates due to the probabilistic transmission at QPC$_\text{R}$. This does not have an impact on the average charge transmitted along each of the classical paths; in contrast it allows to extract which-path information from the fluctuations in the transmitted charge.
The question whether one particle arrives in each reservoir \textit{on average} or whether it is indeed \textit{exactly one} particle in each period, can ultimately be clarified by considering the noise, which we present in Sec.~\ref{sec_noise}. 

Also here, the case where the other type of particles is emitted from the SPSs (namely a hole both from SPS$_\text{A}$ and from SPS$_\text{B}$) leads to a phase difference with respect to the case of two emitted electrons, yielding a finite current in the reservoirs also when considering the total current of one full period, if only $\Delta\tau$ or $\Delta t^{ii}_\text{u}$ are different from zero.

The full general expressions for the charge current in the case of collision and absorption are given in Appendix~\ref{app_analytic_charge}.

\subsection{Energy current}\label{sec_energy_current}

The results of the last section show the impact of absorptions and collisions on the charge current and how they can be explained either based  on the structure of the spectral current or on the occurrence of two-particle effects. Both interpretations are clearly related to the energetic properties of the contributing current pulses, motivating the following discussion of the energy currents detected at the output of the MZI.

In the case where a particle emitted by SPS$_\text{A}$ can possibly be absorbed at SPS$_\text{B}$, the energy current in reservoir~4 is given by
\begin{widetext}
\begin{eqnarray}
&&\bar{J}^\text{eh} =  \frac{\hbar}{2\sigma_\text{A}\mathcal{T}}\left(R_\text{L}R_\text{R}+T_\text{L}T_\text{R}\right)+
\frac{\hbar}{2\sigma_\text{B}\mathcal{T}}T_\text{R}
-T_\text{L}T_\text{R}\left(\frac{\hbar}{2\sigma_\text{A}\mathcal{T}}+\frac{\hbar}{2\sigma_\text{B}\mathcal{T}}\right)
\frac{4\sigma_\text{A}\sigma_\text{B}}{{\Delta t_\text{d}^\text{eh}}^2+\left(\sigma_\text{A}+\sigma_\text{B}\right)^2}\\
&&-2\gamma\frac{\hbar}{2\sigma_\text{A}\mathcal{T}}\mathbb{R}\mathrm{e}\left\{
e^{-i\Phi}\left(\frac{-2i\sigma_\text{A}}{\Delta\tau-2i\sigma_\text{A}}\right)^2\left(1-\frac{2i\sigma_\text{B}}{\Delta t^\text{eh}_\text{d}+i\left(\sigma_\text{A}+\sigma_\text{B}\right)}\right)\right\}\nonumber. 
\end{eqnarray}
\end{widetext}
We see that the synchronisation of the two particle sources affects both the classical as well as the interference part of the energy current. Let us start by considering the classical contribution: while the emission of independent electrons and holes leads to the emission of the same amount of energy related to the width of the current pulse, $\hbar/2\sigma_\text{A/B}$, the absorption of a particle (which can occur when the particle emitted from SPS$_\text{A}$ takes the lower MZI path) leads to an \textit{annihilation} not only of the charge but also of the energy current. The classical part of the energy current thus reduces to  $R_\text{L}R_\text{R}\hbar/2\sigma_\text{A}+ R_\text{L}T_\text{R}\hbar/2\sigma_\text{B}$ in the case of absorption in the lower arm, namely when $\Delta t^\text{eh}_\text{d}=0$ and $\sigma_\text{A}=\sigma_\text{B}$.
 
In the same way we see that the interference is suppressed under the condition, $\Delta t_\text{d}^\text{eh}=0$ and $\sigma_\text{A}=\sigma_\text{B}$, because if the particle is absorbed along the lower path also the energy going along with it does not fluctuate any more at the output and the same coexisting interpretations as for the charge current can possibly be employed, based on the wave and the particle nature of the injected signal. Indeed, we find that the effect of the collision is the suppression of the factor $(-2i\sigma_\text{A})^2/(\Delta\tau-2i\sigma_\text{A})^2$, which was found to be typical for the energy current in the interferometer, see Eq.~(\ref{eq_IE_MZIA}).
The energy current in the case of absorption is shown in Fig.~\ref{fig_charge_energy}~e.) as compared to the case of an MZI with a single working source shown in Fig.~\ref{fig_charge_energy}~d.). Results for the absorption of a hole, namely the synchronised emission of a hole from SPS$_\text{A}$ and an electron from SPS$_\text{B}$ are given in Appendix~\ref{app_analytic_energy}.

Instead, the energy current in the regime where particles of the same type are injected from the two SPSs  such that they arrive in the detector in the same half period is given by
\begin{widetext}
\begin{eqnarray}
\bar{J}^\text{ee}  & = &   \frac{\hbar}{2\sigma_\text{A}\mathcal{T}}\left(R_\text{L}R_\text{R}+T_\text{L}T_\text{R}\right)+
\frac{\hbar}{2\sigma_\text{B}\mathcal{T}}T_\text{R}+
T_\text{L}T_\text{R}\left(\frac{\hbar}{2\sigma_\text{A}\mathcal{T}}+\frac{\hbar}{2\sigma_\text{B}\mathcal{T}}\right)
\frac{4\sigma_\text{A}\sigma_\text{B}}{{\Delta t_\text{d}^\text{ee}}^2+\left(\sigma_\text{A}+\sigma_\text{B}\right)^2}\nonumber\\
&&-2\gamma\frac{\hbar}{2\sigma_\text{A}\mathcal{T}}\mathbb{R}\mathrm{e}\left\{
e^{-i\Phi}\left(\frac{-2i\sigma_\text{A}}{\Delta\tau-2i\sigma_\text{A}}\right)^2\left(1-\frac{-2i\sigma_\text{B}}{\Delta t^\text{ee}_\text{u}-i\left(\sigma_\text{A}+\sigma_\text{B}\right)}\right)\right\}\nonumber\\
&&+2\gamma\frac{\hbar}{2\sigma_\text{B}\mathcal{T}}\mathbb{R}\mathrm{e}\left\{e^{-i\Phi}\frac{-2i\sigma_\text{A}}{\Delta\tau-2i\sigma_\text{A}}\left(\frac{-2i\sigma_\text{B}}{\Delta t^\text{ee}_\text{u}-i\left(\sigma_\text{A}+\sigma_\text{B}\right)}\right)^2
\right\} .
\end{eqnarray} 
\end{widetext}
Also here we show the electronic contribution only; the general expression is given in Appendix~\ref{app_analytic_energy}.
The classical part of the energy current shows an \textit{enhancement} when a particle from SPS$_\text{B}$ is emitted \textit{on top} of a particle emitted from SPS$_\text{A}$ travelling along the lower arm, since the two particles can not occupy the same energy state, due to fermion statistics. This enhancement occurs hence under the condition $\Delta t_\text{d}^\text{ee}=0$ and $\sigma_\text{A}=\sigma_\text{B}\equiv\sigma$ and leads to the classical energy current $\left(R_\text{L}+\boldsymbol{4}T_\text{L}T_\text{R}\right)\hbar/2\sigma$. In contrast, the interference part of the heat current is not affected by this event.

However, like in the case of the charge current, the interference contribution to the heat current is sensitive to possible collisions at the interferometer output taking place if $\Delta t _\text{u}^\text{ee}\approx0$. The interference term contains two contributions: the first is suppressed when the two emitted particles can collide at QPC$_\text{R}$ and one could be tempted to associate it to the corresponding amount of energy of the colliding particles. However, there is an additional term which appears in the vicinity of the collision condition, which stems from the additional oscillations of the spectral current related to the energy scale which can be associated to the time-scale of the particle emission synchronisation, see Eq.~(\ref{eq_spec_coll}).

Intriguingly, the energy current for two particles of the same kind hence behaves rather differently from the charge current: it has features both at the condition $\Delta t_\text{d}^\text{ee}=0$ (classical part) and at the condition $\Delta t_\text{u}^\text{ee}=0$ (interference part) and the interference effects in the energy current do \textbf{not} get suppressed under the collision condition (neither at QPC$_\text{R}$ nor at SPS$_\text{B}$). The collision at QPC$_\text{R}$ rather introduces a phase shift only, which can be seen in Fig.~\ref{fig_charge_energy}~f.). This behaviour has the following important implications.

The continued existence of the interference in the energy current in the case of possible collisions at QPC$_\text{R}$ can obviously not be explained within one consistent particle picture, as it was done for the suppression of interference due to collisions in the charge current. Indeed, when particles can collide at QPC$_\text{R}$, fluctuations in the charge are suppressed while they persist in the energy. Hence, if a particle picture could be used  then it would lead to an apparent \textit{separation} of energy and charge of the particles, namely interference occurring in the energy current while the charge current is flux-independent. This ``paradox" in the particle-interpretation of the energy-charge separation as well as its alternative description by quantum interference has recently been debated for spin-particle~\cite{Aharonov13} and polarisation-particle~\cite{Denkmayr14} separation under the name ``quantum Cheshire cat".~\cite{Correa14,Stuckey14}
 
Finally, we notice that the enhancement of the energy current when collisions at SPS$_\text{B}$ can occur could be considered as a which-path information. It however turns out that this does not influence the interference pattern neither in the charge current nor in the heat current.  Consequently, we find that the coexistence of the interpretations of interference suppression due to phase averages and due to multi-particle effects is to be questioned when energy currents are taken into account.

\section{Two-particle effects from the noise}\label{sec_noise}

In order to better understand true two-particle effects, it is useful to consider the current noise that occurs in the cases studied in the previous sections. Indeed, the noise carries clear signatures of collisions of particles, as it was shown theoretically~\cite{Olkhovskaya08,Feve08,Jonckheere12} as well as experimentally~\cite{Bocquillon12,Dubois13} for the case of the two-particle collider. The collision of particles with the same energy at a beamsplitter leads to a full suppression of the partition noise, since  the two colliding particles are not allowed to enter the same outgoing channel due to fermion statistics.  Equally, the full suppression of the noise in the case of particle absorption in a two-sources setup without an MZI has previously been calculated.~\cite{Moskalets09}

\subsection{Noise of an MZI with one source}

We start by considering the current noise produced by the setup, when SPS$_\text{B}$ is switched off and particles are injected into the MZI by SPS$_\text{A}$, only. The current noise, for the half period in which an electron emitted from SPS$_\text{A}$ arrives at the MZI outputs, can then be written as
\begin{eqnarray}\label{eq_noise_MZIA}
\frac{\mathcal{P}^\text{e}_\text{MZI,A}}{-e^2/\mathcal{T}} & = & T_\mathrm{R}R_\mathrm{R}+T_\mathrm{L}R_\mathrm{L}-4\gamma^2\\
&+& 2\gamma\left(T_\text{L}-R_\text{L}\right)\left(T_\text{R}-R_\text{R}\right)\mathbb{R}\mathrm{e}\left\{
e^{-i\Phi}\frac{-2i\sigma_\text{A}}{\Delta\tau-2i\sigma_\text{A}}\right\}\nonumber\\
&-&\left(2\gamma\mathbb{R}\mathrm{e}\left\{e^{-i\Phi}\frac{-2i\sigma_\text{A}}{\Delta\tau-2i\sigma_\text{A}}\right\}\right)^2
\nonumber\ .
\end{eqnarray}
A similar expression is found for the hole contribution; see the full expression in Appendix~\ref{app_analytic_noise}. Due to the product of current operators contributing to the noise, we here get contributions for the first as well as the second harmonic in the magnetic flux. Since only single particles are emitted into the interferometer per half period it is quite intuitive that we should be able to understand the noise as a simple product of currents. More precisely, it should be proportional to a product of transmission probabilities to the contacts at which the two currents are detected. 

In order to show that, we consider the charge current in the detector, see Eq.~(\ref{eq_curr_MZI}), and rewrite it in terms of effective transmission probabilities, $T^\text{eff,e}_{41}$ and $T^\text{eff,h}_{41}$, for electrons and holes, $\bar{I}_\text{MZI,A}=-e\left(n_\text{A}^\text{e}T^\text{eff,e}_{41}+n_\text{A}^\text{h}T^\text{eff,h}_{41}\right)/\mathcal{T}$, with
\begin{eqnarray*}
T^\text{eff,e}_{41} & = & R_\mathrm{L} R_\mathrm{R}+T_\mathrm{L} T_\mathrm{R} -2\gamma\mathbb{R}\mathrm{e}
\left\{e^{-i\Phi}\frac{-2i\sigma_\mathrm{A}}{\Delta\tau-2i\sigma_\text{A}}\right\}\\
T^\text{eff,h}_{41} & = &-R_\mathrm{L} R_\mathrm{R}-T_\mathrm{L} T_\mathrm{R} +2\gamma\mathbb{R}\mathrm{e}\left\{e^{-i\Phi}\frac{2i\sigma_\mathrm{A}}{\Delta\tau+2i\sigma_\text{A}}\right\}\ .
\end{eqnarray*}
Extracting in an equivalent manner effective transmission probabilities, $T^\text{eff,e}_{31}$ and $T^\text{eff,h}_{31}$, from the current in contact 3,  we are indeed able to show that the noise of the MZI with a single source can simply be written as
\begin{equation}
\mathcal{P}_\text{MZI,A} = -\frac{e^2}{\mathcal{T}}\left[n_\text{A}^\text{e}T_{41}^\text{eff,e}T_{31}^\text{eff,e}+ n_\text{A}^\text{h}T_{41}^\text{eff,h}T_{31}^\text{eff,h}  \right]\ .\label{eq_product}
\end{equation}
This product form of the noise, shown in Eq.~(\ref{eq_product}), is clearly not expected to hold in the case where two particles are injected into the interferometer from different sources and \textit{two-particle effects} will hence contribute to the noise. In order to better understand the impact of two-particle effects, as discussed in the following Sec.~\ref{sec_noise_2}, the following interpretation of the classical part of the noise, given in Eq.~(\ref{eq_noise_MZIA}), turns out to be useful. 
The classical part $T_\mathrm{R}R_\mathrm{R}+T_\mathrm{L}R_\mathrm{L}-4\gamma^2=(R_\mathrm{L} R_\mathrm{R}+T_\mathrm{L} T_\mathrm{R})(R_\mathrm{L} T_\mathrm{R}+T_\mathrm{L} R_\mathrm{R})$, stemming from the product of the classical parts of the effective  transmission probabilities, results in the partition noise of the left and the right QPC, $T_\mathrm{L}R_\mathrm{L}$ and $T_\mathrm{R}R_\mathrm{R}$, and a mixed contribution, $-4\gamma^2$. Furthermore this can be rewritten as $T_\mathrm{R}R_\mathrm{R}+T_\mathrm{L}R_\mathrm{L}-4\gamma^2=T_\text{R}R_\text{R}+T_\text{L}R_\text{L}\left(T_\text{R}-R_\text{R}\right)^2$. It means that the classical part of the noise is given by the partition noise of QPC$_\text{R}$, $T_\text{R}R_\text{R}$, on one hand, and the partition noise of QPC$_\text{L}$ in the presence of QPC$_\text{R}$, $T_\text{L}R_\text{L}\left(T_\text{R}-R_\text{R}\right)^2$, on the other hand. The latter shows that, in the absence of interference,  QPC$_\text{L}$ only produces partition noise if QPC$_\text{R}$ is not symmetric. Indeed, if QPC$_\text{R}$ was symmetric, the probability of particles from SPS$_\text{A}$ to be scattered into the reservoirs~3 and 4 was one half each, independently of the transmission probability of QPC$_\text{L}$, and the partition noise of the latter would thus be invisible.

\subsection{Noise of an MZI with two sources}\label{sec_noise_2}
In the following, we will consider the impact of two-particle effects (absorption and quantum exchange effects) on the charge current noise. 
Let us again start to consider the case where possible absorptions might occur. This is the situation, where indeed the interpretation based on an averaging effect of the spectral current as well as the interpretation based on the absorption of particles, carrying charge and energy, could coexist to explain the occurrence or absence of interference effects even when considering energy currents. In that case the charge-current noise is given by 
\begin{widetext}
\begin{eqnarray}\label{eq_noise_abs}
&&\frac{\mathcal{P}^\text{eh}}{-e^2/\mathcal{T}}   =  R_\mathrm{L}T_\mathrm{L}-4\gamma^2+2 R_\mathrm{L}T_\mathrm{R}R_\mathrm{R}+2T_\mathrm{L}T_\mathrm{R}R_\mathrm{R}\left(1-\frac{4\sigma_\text{A}\sigma_\text{B}}{{\Delta t_\text{d}^\text{eh}}^2+\left(\sigma_\text{A}+\sigma_\text{B}\right)^2}\right)+\\
&& 2\gamma\left(T_\text{L}-R_\text{L}\right)\left(T_\text{R}-R_\text{R}\right)\mathbb{R}\mathrm{e}\left\{
e^{-i\Phi}\frac{-2i\sigma_\text{A}}{\Delta\tau-2i\sigma_\text{A}}
\frac{\Delta t_\text{d}^\text{eh}+i\left(\sigma_\text{A}-\sigma_\text{B}\right)}{\Delta t_\text{d}^\text{eh}+i\left(\sigma_\text{A}+\sigma_\text{B}\right)}
\right\}
-\left(2\gamma\mathbb{R}\mathrm{e}\left\{e^{-i\Phi}\frac{-2i\sigma_\text{A}}{\Delta\tau-2i\sigma_\text{A}}\frac{\Delta t_\text{d}^\text{eh}+i\left(\sigma_\text{A}-\sigma_\text{B}\right)}{\Delta t_\text{d}^\text{eh}+i\left(\sigma_\text{A}+\sigma_\text{B}\right)}
\right\}\right)^2\nonumber .
\end{eqnarray}
\end{widetext}
For the MZI with two sources, we again drop the subscript for the amount of working sources and the presence of the MZI. The classical part of the noise, shown in the first line of Eq.~(\ref{eq_noise_abs}), is partly suppressed by the absorptions. In particular, if the particle from SPS$_\text{A}$ took the lower arm of the interferometer with probability $T_\text{L}$ and could hence get absorbed,  the partition noise at the right barrier created by particles coming from SPS$_\text{A}$ and the opposite type of particle coming from SPS$_\text{B}$, $2T_\text{R}R_\text{R}$, is fully suppressed. 
What is then left from the classical part of the noise is given by $R_\text{L}T_\text{L}-4\gamma^2+2R_\text{L}T_\text{R}R_\text{R}=2R_\text{L}T_\text{R}R_\text{R}+T_\text{L}R_\text{L}\left(T_\text{R}-R_\text{R}\right)^2$. It equals the partition noise of the two particles at  QPC$_\text{R}$ if the particle from SPS$_\text{A}$ took the upper arm, $2R_\text{L}T_\text{R}R_\text{R}$, and the additional noise of the particle from SPS$_\text{A}$ at  QPC$_\text{L}$ in the presence of QPC$_\text{R}$, which can obviously not get affected by the absorptions happening behind QPC$_\text{L}$, $T_\text{L}R_\text{L}\left(T_\text{R}-R_\text{R}\right)^2$. 
Also the interference part of the noise gets fully suppressed by the factor $\frac{\Delta t_\text{d}^\text{eh}+i\left(\sigma_\text{A}-\sigma_\text{B}\right)}{\Delta t_\text{d}^\text{eh}+i\left(\sigma_\text{A}+\sigma_\text{B}\right)}$, in the case of absorptions.
The result for the noise thus fully confirms that the absorption condition leads to a suppression of fluctuations at QPC$_\text{R}$, yielding information on the path that a particle emitted from SPS$_\text{A}$ took in the MZI.

Finally, we consider the case where an electron emitted each from SPS$_\text{A}$ and SPS$_\text{B}$ can reach the reservoirs in the same half period of the source operation. The charge-current noise takes the form
\begin{widetext}
\begin{eqnarray}
&&\frac{\mathcal{P}^\text{ee}}{-e^2/\mathcal{T}}   =  R_\mathrm{L}T_\mathrm{L}-4\gamma^2+2 T_\mathrm{L}T_\mathrm{R}R_\mathrm{R}+2R_\mathrm{L}T_\mathrm{R}R_\mathrm{R}\left(1-\frac{4\sigma_\text{A}\sigma_\text{B}}{{\Delta t_\text{u}^\text{ee}}^2+\left(\sigma_\text{A}+\sigma_\text{B}\right)^2}\right)\\
&& +2\gamma\left(T_\text{L}-R_\text{L}\right)\left(T_\text{R}-R_\text{R}\right)\mathbb{R}\mathrm{e}\left\{
e^{-i\Phi}\frac{-2i\sigma_\text{A}}{\Delta\tau-2i\sigma_\text{A}}
\frac{\Delta t_\text{u}^\text{ee}-i\left(\sigma_\text{A}-\sigma_\text{B}\right)}{\Delta t_\text{u}^\text{ee}-i\left(\sigma_\text{A}+\sigma_\text{B}\right)}
\right\}
-\left(2\gamma\mathbb{R}\mathrm{e}\left\{e^{-i\Phi}\frac{-2i\sigma_\text{A}}{\Delta\tau-2i\sigma_\text{A}}\frac{\Delta t_\text{u}^\text{ee}-i\left(\sigma_\text{A}-\sigma_\text{B}\right)}{\Delta t_\text{u}^\text{ee}-i\left(\sigma_\text{A}+\sigma_\text{B}\right)}
\right\}\right)^2\nonumber .
\end{eqnarray}
\end{widetext}
Equivalently to the absorption case, the behaviour of the charge-current noise corroborates the interpretation of the suppression of interference effects in the charge current based on two-particle collisions. Indeed, only when the collision condition at QPC$_\text{R}$ is fulfilled, the classical part of the noise gets suppressed by the contributions stemming from the partition at QPC$_\text{R}$, when the particle took the upper arm, allowing for collisions at the output of the MZI. The remaining classical noise is then given by $2T_\text{L}T_\text{R}R_\text{R}+T_\text{L}R_\text{L}\left(T_\text{R}-R_\text{R}\right)^2$. At the same time also a full suppression of the interference part of the charge-current noise is found.

Again, the results for the absorption of holes by electrons emitted from SPS$_\text{B}$ and the collision of holes at QPC$_\text{R}$ are shown in the Appendix~\ref{app_analytic_noise}.

\section{Conclusions}
In this work, we studied the charge current and charge-current noise as well as the spectral current and the energy current in an MZI which could be fed by either one or two single-particle sources. When the MZI is fed by only one source, SPS$_\text{A}$, interference effects occur in all four quantities. They are shown to be strongly influenced by the time-scale $\Delta\tau$ stemming from the detuning of the MZI.  
More precisely, the detuning renders the interference contribution to the transmission of the MZI energy-dependent. At finite detuning, this results (1) in a phase shift between the charge and energy current and (2) in a finite dc charge current at each of the MZI outputs, even though the amount of injected electrons and holes is equal. We furthermore show that the  suppression of interference in charge and energy currents for large detuning, $\Delta\tau\gg\sigma_\text{A}$, can be interpreted both as an averaging effect of the interference features occurring in the spectral currents (which represent the plane wave contributions of the injected signals) as well as through the particle-like properties of the injected signal, namely by the limited single-particle coherence length.

In a second step, we investigate the impact of the synchronisation of two SPSs, one of them placed in the centre of the lower interferometer arm, on the quantum-interference effects. Also the synchronisation of the two sources is shown to introduce new relevant time-scales which are related to the absorption or collision of particles at different places in the MZI setup. These new time-scales lead to a suppression of the interference in the spectral current when the sources are tuned to allow for absorptions of particles, or even to the occurrence of additional energy-dependent oscillations when the possibility of collisions of particles of the same type is given. As a result of the occurrence of these new time-scales manifestations of two-particle effects are already visible in the dc charge current.

The absorption of particles at SPS$_\text{B}$, as well as the collision of particles at QPC$_\text{R}$ lead to a suppression of interference in the charge current. Our paper demonstrated that this can be interpreted in two different manners: (1) the suppression of interference can be understood as the result of an averaging of the magnetic-flux dependent contributions of the spectral current. It can on the other hand (2) be explained by the possibility of extracting which-path information from reduced fluctuations due to two-particle effects (absorption and quantum exchange effects). Our investigation of the noise properties corroborates the possibility of a particle-interpretation of the interference suppression by showing that the absorption and collision of particles indeed leads to a specific reduction of fluctuations. 
However, this work also shows that the particle-interpretation does not hold in the case of collisions, whenever the behaviour of the energy current is considered. We show that the energy current behaves fundamentally different from the charge current of electrons and holes displaying signatures of interference when the charge current does not. 

\acknowledgments
We thank Gwendal F\`eve and Patrick Hofer for useful comments on the manuscript. J.~S and M.~M. are grateful for the hospitality at the University of Geneva where part of this work was done.
We acknowledge financial support from the Ministry of Innovation NRW, Germany. Furthermore, financial support from the Excellence Initiative of the
German Federal and State Governments (J.~S. and F.~B.), and from the Knut and Alice Wallenberg foundation through the Wallenberg Academy Fellows program (J.~S.) is acknowledged.

\appendix

\begin{widetext}

\section{Scattering matrices of the MZI with two single-particle sources}\label{app_Smatrix}
In the regime  in which the SPSs are adiabatically driven, the total dynamical scattering matrix for electrons/holes  to be scattered from reservoir $\beta$ to reservoir $\alpha$ of the MZI, fed by  the two sources as described in Section~\ref{sec_model}, contains the following matrix elements
\begin{subequations}
\begin{eqnarray}
S_{\text{in,}41} (t,E)&=&   S_\text{A}(t-\tau_\text{u}) r_\text{L} e^{i\phi_\text{u}(E)}r_\text{R} + S_\text{A}(t-\tau_\text{d})   t_\text{L} S_\text{B}(t-\frac{\tau_\text{d}}{2}) e^{i\phi_\text{d}(E)} t_ \text{R} \\
S_{\text{in,}42}(t,E) &=& t_\text{L} e^{i\phi_\text{u}(E)} r_\text{R}+ r_\text{L} S_\text{B}(t-\frac{\tau_\text{d}}{2}) e^{i\phi_\text{d}(E)}  t_\text{R}  \\
S_{\text{in,}31}(t,E) &=& S_\text{A}(t-\tau_\text{u}) r_\text{L}  e^{i\phi_\text{u}(E)} t_\text{R} + S_\text{A}(t-\tau_\text{d}) t_\text{L}S_\text{B}(t-\frac{\tau_\text{d}}{2})   e^{i\phi_\text{d}(E)} r_\text{R} \\
S_{\text{in,}32}(t,E)&=& t_\text{L}e^{i\phi_\text{u}(E)} t_\text{R} +   r_\text{L}S_\text{B}(t-\frac{\tau_\text{d}}{2}) e^{i\phi_\text{d}(E)}r_\text{R}
\end{eqnarray}
\end{subequations}
All other matrix elements have no relevance for the quantities studied in this paper. Similar expressions are found for the corresponding elements of  $S_{\text{out,}\alpha\beta}(E,t)$.

\section{Synchronized two-particle emission - expressions for the spectral, charge and energy current}\label{app_analytic}

\subsection{Spectral current}\label{app_analytic_spec}

In Section~\ref{sec_2_spectral} we present the spectral currents detected at the output of the MZI when both SPSs are working, leading to the collision of (or the absorption of) electrons. Here, we complement this discussion by presenting the analytic results for 
the spectral current in the case where a hole emitted from SPS$_\text{A}$ encounters an electron emitted from SPS$_\text{B}$
\begin{eqnarray}
i^\text{he}(E,\Phi) & = & 
 R_\mathrm{L}R_\mathrm{R}i_\mathrm{A}^\mathrm{h}(E)+R_\mathrm{L}T_\mathrm{R}i_\mathrm{B}^\mathrm{e}(E)
+T_\mathrm{L}T_\mathrm{R}
		\left(
i_\mathrm{B}^\mathrm{e}(E)+i_\mathrm{A}^\mathrm{h}(E)
	\right)\left(1-\frac{4\sigma_\mathrm{A}\sigma_\mathrm{B}}
			{{\Delta t_\mathrm{d}^\mathrm{he}}^2+\left(\sigma_\mathrm{A}+\sigma_\mathrm{B}\right)^2}\right)
	\\
&& -2\gamma i_\mathrm{A}^\mathrm{h}(E)\mathbb{R}\mathrm{e}\left\{
e^{-i\Phi}e^{-iE\Delta\tau/\hbar}
\left(1-\frac{
			-2i\sigma_\mathrm{B}}
			{\Delta t_\mathrm{d}^\mathrm{he}-i\left(\sigma_\mathrm{A}+\sigma_\mathrm{B}\right)}
\right)
\right\}
\nonumber.
\end{eqnarray}
Furthermore, we find for the hole part of the spectral current in the case of possible collision of holes
\begin{eqnarray}
 && i^\text{hh}(E,\Phi)  =  R_\mathrm{L}R_\mathrm{R}i_\mathrm{A}^\mathrm{h}(E)
 +T_\mathrm{L}T_\mathrm{R}i_\mathrm{A}^\mathrm{h}(E)
\mathbb{R}\mathrm{e}\left\{
 1+
 \frac{4\sigma_\mathrm{A}\sigma_\mathrm{B}}{{\Delta t_\mathrm{d}^\mathrm{hh}}^2+(\sigma_\mathrm{A}-\sigma_\mathrm{B})^2}
+ 2i\sigma_\mathrm{B}\frac{\Delta t_\mathrm{d}^\mathrm{hh}+i(\sigma_\mathrm{A}+\sigma_\mathrm{B})}{{\Delta t_\mathrm{d}^\mathrm{hh}}^2+(\sigma_\mathrm{A}-\sigma_\mathrm{B})^2}e^{-iE\left(\Delta t_\mathrm{d}^\mathrm{hh}-i\left(\sigma_\mathrm{A}-\sigma_\mathrm{B}\right)\right)/\hbar}\right\}\nonumber\\
&& +R_\mathrm{L}T_\mathrm{R}i_\mathrm{B}^\mathrm{h}(E)+T_\mathrm{L}T_\mathrm{R}i_\mathrm{B}^\mathrm{h}(E)
\mathbb{R}\mathrm{e}\left\{
 1+
 \frac{4\sigma_\mathrm{A}\sigma_\mathrm{B}}{{\Delta t_\mathrm{d}^\mathrm{hh}}^2+(\sigma_\mathrm{A}-\sigma_\mathrm{B})^2}
+ 2i\sigma_\mathrm{A}\frac{\Delta t_\mathrm{d}^\mathrm{hh}+i(\sigma_\mathrm{A}+\sigma_\mathrm{B})}{{\Delta t_\mathrm{d}^\mathrm{hh}}^2+(\sigma_\mathrm{A}-\sigma_\mathrm{B})^2}e^{-iE\left(\Delta t_\mathrm{d}^\mathrm{hh}-i\left(\sigma_\mathrm{B}-\sigma_\mathrm{A}\right)\right)/\hbar}\right\}\nonumber\\
&&-2\gamma i_\mathrm{A}^\mathrm{h}(E)\mathbb{R}\mathrm{e}\left\{
e^{-i\Phi}e^{-iE\Delta\tau/\hbar}
\left[1-\frac{2i\sigma_\mathrm{B}}{\Delta t_\mathrm{d}^\mathrm{hh}-i\left(\sigma_\mathrm{A}-\sigma_\mathrm{B}\right)}
\left(1-e^{-iE\left(\Delta t_\mathrm{d}^\mathrm{hh}-i\left(\sigma_\mathrm{A}-\sigma_\mathrm{B}\right)\right)/\hbar}\right)
\right]
\right\} .
\end{eqnarray}
In order to find the limit in which either SPS$_\text{A}$ of SPS$_\text{B}$ is switched off, it is enough to set $\sigma_\text{A}\rightarrow0$ (respectively, $\sigma_\text{B}\rightarrow0$). The same applies for Eqs.~(\ref{eq_spec_abs}) and (\ref{eq_spec_coll}).
\subsection{Charge current}\label{app_analytic_charge}

All expressions  for the time-averaged charge current given in the main text in the regime where particles of opposite type arrive in the detector from the two SPSs can be obtained from the general expression
\begin{eqnarray}
\frac{\bar{I}^\text{eh+he}}{e/\mathcal{T}} & = & R_\text{L}R_\text{R}\left(n^\text{h}_\text{A}-n^\text{e}_\text{A}\right) + 
R_\text{L}T_\text{R}\left(n^\text{h}_\text{B}-n^\text{e}_\text{B}\right)
+ T_\text{L}T_\text{R}\frac{\Delta t_\text{d}^2+\left(\sigma_\text{A}-\sigma_\text{B}\right)^2}{\Delta t_\text{d}^2+\left(\sigma_\text{A}+\sigma_\text{B}\right)^2}\left(n^\text{h}_\text{A}-n^\text{e}_\text{A}+n^\text{h}_\text{B}-n^\text{e}_\text{B}\right)\\
&-&2\gamma\mathbb{R}\mathrm{e}\left\{e^{-i\Phi}\left(
n^\text{h}_\text{A}\frac{2i\sigma_\text{A}}{\Delta\tau+2i\sigma_\text{A}}
\left(1-n^\text{e}_\text{B}\frac{-2i\sigma_\text{B}}{\Delta t_\text{d}-i\left(\sigma_\text{A}+\sigma_\text{B}\right)}\right)
-n^\text{e}_\text{A}\frac{-2i\sigma_\text{A}}{\Delta\tau-2i\sigma_\text{A}}
\left(1-n^\text{h}_\text{B}\frac{2i\sigma_\text{B}}{\Delta t_\text{d}+i\left(\sigma_\text{A}+\sigma_\text{B}\right)}\right)
\right)\right\}\nonumber
\end{eqnarray}
by setting the respective particle numbers $n^i_k=0,1$.  Here, we assume that the time difference $\Delta t_\text{d}^\text{eh}=\Delta t_\text{d}^\text{he}\equiv\Delta t_\text{d}$ is equal for electrons and holes. However, different collision conditions $\Delta t^{ij}_\text{d}$ can be obtained straightforwardly by adjusting them for each contribution $n^i_k$.
The result for the MZI with a single SPS$_\text{A}$ is found by setting $n^\text{e}_\text{B}=n^\text{h}_\text{B}=0$. Also $\sigma_\text{B}$ equals zero if SPS$_\text{B}$ is switched off. 

The general expression for the charge current  in the regime where particles of the same type arrive in the detector from both SPSs is
\begin{eqnarray}
\frac{\bar{I}^{\text{ee}+\text{hh}}}{e/\mathcal{T}} & = &  R_\text{L}R_\text{R}\left(n^\text{h}_\text{A}-n^\text{e}_\text{A}\right) + 
R_\text{L}T_\text{R}\left(n^\text{h}_\text{B}-n^\text{e}_\text{B}\right)
+ T_\text{L}T_\text{R}\frac{\Delta t_\text{d}^2+\left(\sigma_\text{A}+\sigma_\text{B}\right)^2}{\Delta t_\text{d}^2+\left(\sigma_\text{A}-\sigma_\text{B}\right)^2}\left(n^\text{h}_\text{A}+n^\text{h}_\text{B}-n^\text{e}_\text{A}-n^\text{e}_\text{B}\right)\\
&&-2T_\text{L}T_\text{R}\frac{4\sigma_\text{A}\sigma_\text{B}}{\Delta t_\text{d}^2+\left(\sigma_\text{A}-\sigma_\text{B}\right)^2}\left( n^\text{h}_\text{A}n^\text{h}_\text{B}-n^\text{e}_\text{A}n^\text{e}_\text{B}\right)-2\gamma\mathbb{R}\mathrm{e}\left\{e^{-i\Phi}\left(n^\text{h}_\text{A}\frac{2i\sigma_\text{A}}{\Delta\tau+2i\sigma_\text{A}}-n^\text{e}_\text{A}\frac{-2i\sigma_\text{A}}{\Delta\tau-2i\sigma_\text{A}}\right)\right\}\nonumber\\
&& +2\gamma\mathbb{R}\mathrm{e}\left\{e^{-i\Phi}\left(
n^\text{h}_\text{A}n^\text{h}_\text{B}\frac{2i\sigma_\text{A}}{\Delta\tau+2i\sigma_\text{A}}
\frac{2i\sigma_\text{B}}{\Delta t_\text{u}+i\left(\sigma_\text{A}+\sigma_\text{B}\right)}
-n^\text{e}_\text{A}n^\text{e}_\text{B}\frac{-2i\sigma_\text{A}}{\Delta\tau-2i\sigma_\text{A}}
\frac{-2i\sigma_\text{B}}{\Delta t_\text{u}-i\left(\sigma_\text{A}+\sigma_\text{B}\right)}
\right)\right\}
\nonumber .
\end{eqnarray}
Also here we took $\Delta t_\text{d}^\text{ee}=\Delta t_\text{d}^\text{hh}\equiv\Delta t_\text{d}$ and $\Delta t_\text{u}^\text{ee}=\Delta t_\text{u}^\text{hh}\equiv\Delta t_\text{u}$ for simplicity.

\subsection{Energy current}\label{app_analytic_energy}

Similar to the case of the charge current, we only show a part of the different particle contributions to the energy current in the main text. In this appendix we report the full expressions, where the same considerations for the different contributing particles, $n_k^\text{e}$ and $n_k^\text{h}$, and the time differences characterising their synchronised emissions, $\Delta t_\text{u}^{ij}$ and $\Delta t_\text{d}^{ij}$, apply, as it was explained for the charge currents in Appendix~\ref{app_analytic_charge}. 

When the SPSs are tuned such that particles of different type emitted from the two sources arrive at the detector in the same half period and hence absorptions can possibly occur, the general expression for the energy current is
\begin{eqnarray}
\bar{J}^\text{eh+he} & = & \frac{\hbar}{2\sigma_\text{A}\mathcal{T}}\left(n_\text{A}^\text{e}+n_\text{A}^\text{h}\right)\left(R_\text{L}R_\text{R}+T_\text{L}T_\text{R}\right)+
\frac{\hbar}{2\sigma_\text{B}\mathcal{T}}\left(n_\text{B}^\text{e}+n_\text{B}^\text{h}\right)T_\text{R}\\
&&-
T_\text{L}T_\text{R}\left(\frac{\hbar}{2\sigma_\text{A}\mathcal{T}}+\frac{\hbar}{2\sigma_\text{B}\mathcal{T}}\right)\left(n_\text{A}^\text{e}n_\text{B}^\text{h}+n_\text{A}^\text{h}n_\text{B}^\text{e}\right)
\frac{4\sigma_\text{A}\sigma_\text{B}}{\Delta t_\text{d}^2+\left(\sigma_\text{A}+\sigma_\text{B}\right)^2}\nonumber\\
&&-2\gamma\frac{\hbar}{2\sigma_\text{A}\mathcal{T}}\mathbb{R}\mathrm{e}\left\{
e^{-i\Phi}\left[n_\text{A}^\text{e}\left(\frac{-2i\sigma_\text{A}}{\Delta\tau-2i\sigma_\text{A}}\right)^2\left(1-n_\text{B}^\text{h}\frac{2i\sigma_\text{B}}{\Delta t_\text{d}+i\left(\sigma_\text{A}+\sigma_\text{B}\right)}\right)\right.\right.\nonumber\\
&&\left.\left.+n_\text{A}^\text{h}\left(\frac{2i\sigma_\text{A}}{\Delta\tau+2i\sigma_\text{A}}\right)^2\left(1-n_\text{B}^\text{e}\frac{-2i\sigma_\text{B}}{\Delta t_\text{d}-i\left(\sigma_\text{A}+\sigma_\text{B}\right)}\right)\right]
\right\}\nonumber .
\end{eqnarray} 

For the regime in which collisions between particles can occur, we find
\begin{eqnarray}
\bar{J}^\text{ee+hh} & = & \frac{\hbar}{2\sigma_\text{A}\mathcal{T}}\left(n_\text{A}^\text{e}+n_\text{A}^\text{h}\right)\left(R_\text{L}R_\text{R}+T_\text{L}T_\text{R}\right)+
\frac{\hbar}{2\sigma_\text{B}\mathcal{T}}\left(n_\text{B}^\text{e}+n_\text{B}^\text{h}\right)T_\text{R}\\
&&+
T_\text{L}T_\text{R}\left(\frac{\hbar}{2\sigma_\text{A}\mathcal{T}}+\frac{\hbar}{2\sigma_\text{B}\mathcal{T}}\right)\left(n_\text{A}^\text{e}n_\text{B}^\text{e}+n_\text{A}^\text{h}n_\text{B}^\text{h}\right)
\frac{4\sigma_\text{A}\sigma_\text{B}}{\Delta t_\text{d}^2+\left(\sigma_\text{A}+\sigma_\text{B}\right)^2}\nonumber\\
&&-2\gamma\frac{\hbar}{2\sigma_\text{A}\mathcal{T}}\mathbb{R}\mathrm{e}\left\{
e^{-i\Phi}\left[n_\text{A}^\text{e}\left(\frac{-2i\sigma_\text{A}}{\Delta\tau-2i\sigma_\text{A}}\right)^2\left(1-n_\text{B}^\text{e}\frac{-2i\sigma_\text{B}}{\Delta t_\text{u}-i\left(\sigma_\text{A}+\sigma_\text{B}\right)}\right)\right.\right.\nonumber\\
&&\left.\left.+n_\text{A}^\text{h}\left(\frac{2i\sigma_\text{A}}{\Delta\tau+2i\sigma_\text{A}}\right)^2\left(1-n_\text{B}^\text{h}\frac{2i\sigma_\text{B}}{\Delta t_\text{u}+i\left(\sigma_\text{A}+\sigma_\text{B}\right)}\right)\right]
\right\}\nonumber\\
&&+2\gamma\frac{\hbar}{2\sigma_\text{B}\mathcal{T}}\mathbb{R}\mathrm{e}\left\{e^{-i\Phi}\left[
n_\text{A}^\text{e}n_\text{B}^\text{e}
\frac{-2i\sigma_\text{A}}{\Delta\tau-2i\sigma_\text{A}}\left(\frac{-2i\sigma_\text{B}}{\Delta t_\text{u}-i\left(\sigma_\text{A}+\sigma_\text{B}\right)}\right)^2
+n_\text{A}^\text{h}n_\text{B}^\text{h}
\frac{2i\sigma_\text{A}}{\Delta\tau+2i\sigma_\text{A}}\left(\frac{2i\sigma_\text{B}}{\Delta t_\text{u}+i\left(\sigma_\text{A}+\sigma_\text{B}\right)}\right)^2\right]
\right\}.
\nonumber
\end{eqnarray} 

\section{Analytic expressions for the noise}\label{app_analytic_noise}
Finally, we consider the charge-current noise, stemming from the current-current correlator of the currents detected in reservoirs~3 and 4. If SPS$_\text{B}$ is switched off and particles are emitted into the MZI only from SPS$_\text{A}$, the total noise stemming from electrons and holes is given by
\begin{eqnarray}
\frac{\mathcal{P}_\text{MZI,A}}{-e^2/\mathcal{T}} & = & \left(T_\mathrm{R}R_\mathrm{R}+T_\mathrm{L}R_\mathrm{L}-4\gamma^2\right)\left(n_\text{A}^\text{e}+n_\text{A}^\text{h}\right)\\
&& +2\gamma\left(T_\text{L}-R_\text{L}\right)\left(T_\text{R}-R_\text{R}\right)\mathbb{R}\mathrm{e}\left\{
e^{-i\Phi}\left(
n_\text{A}^\text{e}\frac{-2i\sigma_\text{A}}{\Delta\tau-2i\sigma_\text{A}}+n_\text{A}^\text{h}\frac{2i\sigma_\text{A}}{\Delta\tau+2i\sigma_\text{A}}
\right)
\right\}\nonumber\\
&&-n_\text{A}^\text{e}\left(2\gamma\mathbb{R}\mathrm{e}\left\{e^{-i\Phi}\frac{-2i\sigma_\text{A}}{\Delta\tau-2i\sigma_\text{A}}\right\}\right)^2
-n_\text{A}^\text{h}\left(2\gamma\mathbb{R}\mathrm{e}\left\{e^{-i\Phi}\frac{2i\sigma_\text{A}}{\Delta\tau+2i\sigma_\text{A}}\right\}\right)^2
\nonumber .
\end{eqnarray}

The noise for the case of  a possible  absorption of a hole emitted by SPS$_\text{A}$ by an emission of an electron from SPS$_\text{B}$ is given by
\begin{eqnarray}
&&\frac{\mathcal{P}^\text{he}}{-e^2/\mathcal{T}}   =  R_\mathrm{L}T_\mathrm{L}-4\gamma^2+2 R_\mathrm{L}T_\mathrm{R}R_\mathrm{R}+2T_\mathrm{L}T_\mathrm{R}R_\mathrm{R}\left(1-\frac{4\sigma_\text{A}\sigma_\text{B}}{{\Delta t_\text{d}^\text{he}}^2+\left(\sigma_\text{A}+\sigma_\text{B}\right)^2}\right)+\\
&& 2\gamma\left(T_\text{L}-R_\text{L}\right)\left(T_\text{R}-R_\text{R}\right)\mathbb{R}\mathrm{e}\left\{
e^{-i\Phi}\frac{2i\sigma_\text{A}}{\Delta\tau+2i\sigma_\text{A}}
\frac{\Delta t_\text{d}^\text{he}-i\left(\sigma_\text{A}-\sigma_\text{B}\right)}{\Delta t_\text{d}^\text{he}-i\left(\sigma_\text{A}+\sigma_\text{B}\right)}
\right\}\nonumber\\
&&-\left(2\gamma\mathbb{R}\mathrm{e}\left\{e^{-i\Phi}\frac{2i\sigma_\text{A}}{\Delta\tau+2i\sigma_\text{A}}\frac{\Delta t_\text{d}^\text{he}-i\left(\sigma_\text{A}-\sigma_\text{B}\right)}{\Delta t_\text{d}^\text{he}-i\left(\sigma_\text{A}+\sigma_\text{B}\right)}
\right\}\right)^2\nonumber.
\end{eqnarray}
For the noise in the case of the collision of two holes we find
\begin{eqnarray}
&&\frac{\mathcal{P}^\text{hh}}{-e^2/\mathcal{T}}   =  R_\mathrm{L}T_\mathrm{L}-4\gamma^2+2 T_\mathrm{L}T_\mathrm{R}R_\mathrm{R}+2R_\mathrm{L}T_\mathrm{R}R_\mathrm{R}\left(1-\frac{4\sigma_\text{A}\sigma_\text{B}}{{\Delta t_\text{u}^\text{hh}}^2+\left(\sigma_\text{A}+\sigma_\text{B}\right)^2}\right)\\
&& +2\gamma\left(T_\text{L}-R_\text{L}\right)\left(T_\text{R}-R_\text{R}\right)\mathbb{R}\mathrm{e}\left\{
e^{-i\Phi}\frac{2i\sigma_\text{A}}{\Delta\tau+2i\sigma_\text{A}}
\frac{\Delta t_\text{u}^\text{hh}+i\left(\sigma_\text{A}-\sigma_\text{B}\right)}{\Delta t_\text{u}^\text{hh}+i\left(\sigma_\text{A}+\sigma_\text{B}\right)}
\right\} 
\nonumber\\&&
-\left(2\gamma\mathbb{R}\mathrm{e}\left\{e^{-i\Phi}\frac{2i\sigma_\text{A}}{\Delta\tau+2i\sigma_\text{A}}\frac{\Delta t_\text{u}^\text{hh}+i\left(\sigma_\text{A}-\sigma_\text{B}\right)}{\Delta t_\text{u}^\text{hh}+i\left(\sigma_\text{A}+\sigma_\text{B}\right)}
\right\}\right)^2\nonumber\ .
\end{eqnarray}
\end{widetext}


%

\end{document}